\documentclass[12pt,a4paper]{article}
\usepackage{amsmath,amssymb,cite,slashed,tabularx}
\usepackage{subfigure}
\usepackage{graphicx,xcolor}

\allowdisplaybreaks

\numberwithin{equation}{section} 
\def\beq#1\eeq{\begin{align}#1\end{align}}

\usepackage{caption}
\definecolor{BlueViolet}{rgb}{0.2, 0.00, 0.7}
\definecolor{Blue}{rgb}{0.15, 0.00, 0.9}
\usepackage[colorlinks=true,linkcolor=Blue,citecolor=Blue,urlcolor=BlueViolet]{hyperref}

\begin{document}

\begin{titlepage}

\begin{center}

\hfill KEK--TH--2078\\
\hfill TTP18--032\\
\hfill November 2018\\

\vskip .35in

{\Large\bf
 SMEFT top-quark effects on $\Delta F=2$ observables
}

\vskip .4in

{\large
  Motoi Endo$^{\rm (a,b)}$, 
  Teppei Kitahara$^{\rm (c,d,e,f,g)}$,
  Daiki Ueda$^{\rm (b,c)}$
}

\vskip 0.25in

$^{\rm (a)}${\it
Theory Center, IPNS, KEK, Tsukuba, Ibaraki 305-0801, Japan}

\vskip 0.1in

$^{\rm (b)}${\it
The Graduate University of Advanced Studies (Sokendai),\\
Tsukuba, Ibaraki 305-0801, Japan}

\vskip 0.1in

$^{\rm (c)}${\it 
Institute for Theoretical Particle Physics (TTP), Karlsruhe Institute of Technology, Engesserstra{\ss}e 7, D-76128 Karlsruhe, Germany}
 
\vskip 0.1in
 
$^{\rm (d)}${\it
Institute for Nuclear Physics (IKP), Karlsruhe Institute of
Technology, Hermann-von-Helmholtz-Platz 1, D-76344
Eggenstein-Leopoldshafen, Germany} 

\vskip 0.1in

$^{\rm (e)}${\it 
Institute for Advanced Research, Nagoya University,
Furo-cho Chikusa-ku, Nagoya, Aichi, 464-8602 Japan}

\vskip 0.1in

$^{\rm (f)}${\it 
Kobayashi-Maskawa Institute for the Origin of Particles and the
Universe, \\ Nagoya University, Nagoya 464-8602, Japan}

\vskip 0.1in

$^{\rm (g)}${\it 
Physics Department, Technion--Israel Institute of Technology, Haifa 3200003, Israel
}

\end{center}

\vskip .3in

\begin{abstract}

We investigate model independent top-quark corrections to $\Delta F = 2$ processes for the down-type quarks within the framework of the Standard Model Effective Field Theory.
Dimension-six $\Delta F = 1$ operators contribute to them through renormalization group evolutions and matching conditions. 
We provide a complete one-loop matching formula from the top quarks for $\Delta F=2$ transitions.
We also demonstrate these corrections on $\Delta M_{B_s}$ in the left-right symmetric model, which are compared with the conventional calculation.

\vskip .3in


\end{abstract}
\end{titlepage}

\setcounter{page}{1}
\renewcommand{\thefootnote}{\#\arabic{footnote}}
\setcounter{footnote}{0}

\section{Introduction}
\label{sec:introduction}

Flavor-changing neutral currents (FCNCs) are sensitive to physics at high energy scales.
New particles predicted by new physics (NP) models could contribute to FCNCs in addition to the weak boson and the heavy quarks in the Standard Model (SM).

FCNCs are absent at the tree level in the SM and are produced at the loop level.
After the electroweak symmetry breaking (EWSB) and integrating the weak boson and the heavy quarks out, they are represented by higher dimensional operators. 
When the NP scale is much higher than the EWSB scale, the NP contributions to FCNCs are also encoded as the higher dimensional operators in the Standard Model Effective Field Theory (SMEFT)~\cite{Buchmuller:1985jz, Grzadkowski:2010es, Dedes:2017zog} above the EWSB scale. 
Here, the higher dimensional operators are invariant under the SM gauge symmetries, SU(3$)_{C} \times$ SU(2$)_{L} \times$ U(1$)_{Y}$, and all the SM particles, particularly the electroweak bosons ($W,Z,H$) and the top quark ($t$), are dynamical degrees of freedom.
By integrating out $W,Z,H$ and $t$ at the EWSB scale, the SMEFT effective operators are matched to operators in the low energy effective Hamiltonian; they are eventually compared with experimental data, and let us call them the ``low-scale'' effective operators.

In a class of the NP models, both of NP and SM particles appear simultaneously in a loop diagram.
The NP particles are likely to be heavier than the SM ones, since absent discoveries of new particles at the LHC push the NP scale higher than the EWSB one~\cite{LHC}.
Then, the fixed-order perturbative calculations are inappropriate. 
When there is a large mass hierarchy among the particles in a loop diagram, higher order corrections of the perturbation cannot be negligible.
In particular, it is uncertain in which energy scale the model parameters are evaluated.

Corrections of the dynamical top quark\footnote{
Above the EWSB scale, the top quark is one of dynamical degrees of freedom.
Its dynamics provides $\mathcal{O}(y_t^2)$ corrections and operator mixings during renormalization group evolutions.
} to the low-scale effective operators can be relevant. 
This is not surprising because the top quark has a large Yukawa coupling and mass, and it is charged under the SU(3$)_{C}$ symmetry.
In FCNC amplitudes, heavy particle contributions tend to be sizable due to ``the GIM mechanism'' in analogy to the SM case~\cite{Glashow:1970gm}. 
When the GIM suppression is broken by the up-type quark masses in a NP loop diagram analogously to the SM case, the top quark contribution could be dominant, depending on flavor structures of NP couplings. 
Then, the above problem is rephrased as ``in which energy scale the top quark mass (or the top Yukawa coupling) is evaluated.''
This is resolved by means of the renormalization group equations (RGEs). 
It is necessary to solve RGEs in the SMEFT.

In a conventional approach, however, the NP diagrams are matched directly to the low-scale effective operators by integrating out the NP particles and the top quark as well, and the SMEFT effects, i.e., corrections from the dynamical $t$ (and also $W,Z,H$), are discarded.
Such an approximation generates large logarithms $\ln(m_t/M_{\rm NP})$ and then lead to a large scale uncertainty because of a lack of $\mathcal{O}(y_t^2)$ corrections from the SMEFT RGRs.
This scale uncertainty becomes larger, when the NP scale becomes higher than the EWSB scale.
In this paper, we study the SMEFT corrections above the EWSB scale, paying particular attention to the dynamical top quark.\footnote{
  It is straightforward to extend our study to the lighter quarks such as the up- or charm-quarks.
  In this case, however, the one-loop matching formula at the EWSB scale discussed below is irrelevant, and long-distance effects should be taken into account. 
}

We focus on $\Delta F=2$ processes of the down-type quarks, i.e., the $K^0\textrm{--}\overline{K}{}^0$ and $B^0_q\textrm{--}\overline{B}{}^0_q$ $(q=d,s)$ oscillations.
The top quark contributions must be treated carefully.
Above the EWSB scale, there exist $\Delta F=1$ effective operators which subsequently contribute to the low-scale $\Delta F=2$ operators through $W$ and $H$ exchanges.
For instance, 
the $\Delta F = 1$ SMEFT quark--Higgs operators, e.g., $(\bar{q}_i \gamma^{\mu} P_R q_j)(H^{\dag} i \overleftrightarrow{D}_{\mu} H)$, have been studied in Ref.~\cite{Bobeth:2017xry} (see also Ref.~\cite{Endo:2016tnu}).
It is noticed that the one-loop matching corrections at the EWSB can be sizable in a certain class of NP models. 
This is because the corrections are generated with the top-quark mass or Yukawa coupling as well as the CKM matrix, and also because FCNCs are induced at loop levels in many models.
In this paper, we provide a complete one-loop formula for the $\Delta F=1$ contributions to the low-scale $\Delta F=2$ operators with the top Yukawa couplings.\footnote{
  A part of the one-loop matching formula is shown in Ref.~\cite{Aebischer:2015fzz}. 
  We found that its result is inadequate because the left-handed top quark contributions are missing, and thus, inconsistent with the SMEFT RGEs~\cite{private}.
  In addition, the logarithmic scale dependence in Eqs.~(4.24)--(4.26) of the journal version of Ref.~\cite{Aebischer:2015fzz} is inconsistent with that from the RGEs, which are fixed in our result, Eqs.~\eqref{eq:EWSBmatchingLoop1}--\eqref{eq:EWSBmatchingLoop3}.
  The formula related to the SMEFT quark--Higgs operators are given in Ref.~\cite{Bobeth:2017xry} (see also Ref.~\cite{Endo:2016tnu});
  the result is included in this paper.
}
As a demonstration, we study the left-right symmetric models~\cite{Pati:1974yy,Mohapatra:1974hk,Mohapatra:1974gc,Senjanovic:1975rk,Senjanovic:1978ev}, where a new $W$ boson and heavy Higgs bosons induce $\Delta F=1$ effective operators at the NP scale.

\section{Formula}
\label{sec:formula}

In this section, we provide the formula for the SMEFT corrections at the one-loop level which contribute to the $\Delta F=2$ processes of the down-type quarks in a low-energy scale.
In the SMEFT, it is assumed that NP models do not break the electroweak symmetry explicitly. 
After integrating out the heavy NP particles at the NP scale, which is assumed to be much higher than the EWSB scale, $\Delta F = 1$ and $\Delta F = 2$ effects are encoded into higher dimensional operators in the SMEFT, which are defined as~\cite{Grzadkowski:2010es}
\begin{align}
 \mathcal{L}_{\rm eff} = \mathcal{L}_{\textrm{SM}} + \sum_i C_i \mathcal{O}_i,
 \label{eq:SMEFT}
\end{align}
where the first term is the SM Lagrangian at the renormalizable level, and the second term represents the higher dimensional operators.
The dimension-six operators relevant for the low-scale $\Delta F=2$ processes of the down-type quarks are shown as
\begin{align}
 (\mathcal{O}_{qq}^{(1)})_{ijkl} &= 
 (\overline{q}^i \gamma_{\mu} q^j)(\overline{q}^k \gamma_{\mu} q^l),
 \\
 (\mathcal{O}^{(3)}_{qq})_{ijkl} &= 
 (\overline{q}^i\gamma_{\mu}\tau^I q^j)(\overline{q}^k \gamma^{\mu}\tau^I q^l),
 \\
 (\mathcal{O}^{(1)}_{qd})_{ijkl} &= 
 (\overline{q}^i\gamma_{\mu} q^j)(\overline{d}^k \gamma^{\mu} d^l),
 \\
 (\mathcal{O}^{(8)}_{qd})_{ijkl} &= 
 (\overline{q}^i \gamma_{\mu}T^A q^j)(\overline{d}^k \gamma^{\mu}T^A d^l),
 \\
 (\mathcal{O}_{dd})_{ijkl} &= 
 (\overline{d}^i\gamma_{\mu} d^j)(\overline{d}^k \gamma^{\mu} d^l),
 \\
 (\mathcal{O}^{(1)}_{Hq})_{ij} &= 
 (H^\dagger i\overleftrightarrow{D_\mu} H)(\overline{q}^{i} \gamma^{\mu} q^{j}),
 \\
 (\mathcal{O}^{(3)}_{Hq})_{ij} &= 
 (H^\dagger i\overleftrightarrow{D^I_{\mu}} H)(\overline{q}^{i} \gamma^{\mu}\tau^I q^{j}), 
 \\
 (\mathcal{O}_{Hd})_{ij} &= 
 (H^\dagger i\overleftrightarrow{D_{\mu}} H)(\overline{d}^{i} \gamma^{\mu} d^{j}),
\end{align}
with the derivative,
\begin{align}
 H^\dagger\overleftrightarrow{D^I_\mu} H = 
 H^{\dagger} \tau^I D_{\mu} H - \left( D_{\mu} H\right)^{\dagger}  \tau^I H,
\end{align}
where $q$ is the ${\rm SU(2)}_L$ quark doublet, $d$ the right-handed down-type quark, and $T^A$ the ${\rm SU(3)}_C$ generator with quark-flavor indices $i,j,k,l$ and an ${\rm SU(2)}_L$ [${\rm SU(3)}_C$] index $I$ ($A$).
We focus on the top-Yukawa and  QCD interactions.
In addition, the following dimension-six operators have to be included in the analysis:
\begin{align}
 (\mathcal{O}^{ (1)}_{qu})_{ijkl} &= 
 (\overline{q}^i \gamma_{\mu} q^j)(\overline{u}^k \gamma^{ \mu} u^l),
 \\
 (\mathcal{O}^{(8)}_{qu})_{ijkl} &= 
 (\overline{q}^i \gamma_{\mu}T^A q^j)(\overline{u}^k \gamma^{\mu}T^A u^l),
 \\
 (\mathcal{O}_{uu})_{ijkl} &= 
 (\overline{u}^i \gamma_{\mu} u^j)(\overline{u}^k \gamma^{\mu} u^l),
 \\
 (\mathcal{O}^{(1)}_{ud})_{ijkl} &= 
 (\overline{u}^i \gamma_{\mu} u^j)(\overline{d}^k \gamma^{\mu} d^l),
 \\
 (\mathcal{O}^{(8)}_{ud})_{ijkl} &= 
 (\overline{u}^i \gamma_{\mu} T^A u^j)(\overline{d}^k \gamma^{\mu}T^A d^l),
 \\
 (\mathcal{O}_{Hu})_{ij} &= 
 (H^\dagger i\overleftrightarrow{D_{\mu}} H)(\overline{u}^{i} \gamma^{\mu} u^{j}),
 \\
 (\mathcal{O}_{H\Box})_{ij} &= 
 (H^\dagger H) \Box (H^\dagger H),
 \\
 (\mathcal{O}_{HD})_{ij} &= 
 (H^\dagger D_{\mu} H)^* (H^\dagger D^{\mu} H),
\end{align}
where $u$ is the right-handed up-type quark. 
These operators contribute to the $\Delta F=2$ observables through the operator mixings during the RG evolutions and the matching conditions at the EWSB scale (see below). 
Once they are set at the NP scale, the SMEFT RGEs are solved at the one-loop level.
The SMEFT RGEs relevant to the $\Delta F=2$ observables are listed in Appendix~\ref{sec:RGE}.
We keep the anomalous dimension terms which depend on the top Yukawa or QCD couplings. 

The SM heavy degrees of freedom, $W,Z,H$ and $t$, are integrated out at the EWSB scale.
The SMEFT operators are matched to the effective operators in the low-energy scale. 
The low-scale $\Delta F=2$ operators are defined as \cite{Gabbiani:1996hi}
\begin{align}
 \mathcal{H}_{\rm eff}^{\Delta F=2} &=
   (C_1)_{ij} (\overline{d}_i \gamma^{\mu} P_L d_j)(\overline{d}_i \gamma_{\mu} P_L d_j)
 \notag \\ & 
 + (C_2)_{ij} (\overline{d}_i P_L d_j) (\bar{d}_i P_L d_j)
 + (C_3)_{ij} (\overline{d}_i^{\alpha} P_L d_j^{\beta})(\overline{d}_i^{\beta} P_L d_j^{\alpha})
 \notag \\ & 
 + (C_4)_{ij} (\overline{d}_i P_L d_j)(\overline{d}_i P_R d_j)
 + (C_5)_{ij} (\overline{d}_i^{\alpha} P_L d_j^{\beta})(\overline{d}_i^{\beta} P_R d_j^{\alpha})
 \notag \\ &
 + (C^{\prime}_1)_{ij} (\overline{d}_i \gamma^{\mu} P_R d_j)(\overline{d}_i \gamma_{\mu} P_R d_j)
 \notag \\ & 
 + (C^{\prime}_2)_{ij} (\overline{d}_i P_R d_j) (\bar{d}_i P_R d_j)
 + (C^{\prime}_3)_{ij} (\overline{d}_i^{\alpha} P_R d_j^{\beta})(\overline{d}_i^{\beta} P_R d_j^{\alpha}),
 \label{Hamiltonianeff}
\end{align}
where $i,j$ ($i\neq j$) are flavor indices, and $\alpha,\beta$ are color ones.

At the tree level, they are related to the SMEFT operators as
\begin{align}
 (C_1)_{ij}^{\rm tree} &= - \left[ (C^{(1)}_{qq})_{ijij} + (C^{(3)}_{qq})_{ijij} \right], 
 \label{eq:EWSBmatchingTree1} \\
 (C'_1)_{ij}^{\rm tree} &= - (C_{dd})_{ijij},
 \\
 (C_4)_{ij}^{\rm tree} &= (C^{(8)}_{qd})_{ijij},
 \\
 (C_5)_{ij}^{\rm tree} &= 2 (C^{(1)}_{qd})_{ijij} - \frac{1}{N_c} (C^{(8)}_{qd})_{ijij},
 \label{eq:EWSBmatchingTree4}
\end{align}
where the Wilson coefficients in the left-handed side are defined in the low-scale basis, Eq.~\eqref{Hamiltonianeff}, and those in the right-handed side are defined in the SMEFT, Eq.~\eqref{eq:SMEFT}.  
Both of them are evaluated as a weak scale, $\mu=\mu_{W}$.
The other low-scale $\Delta F=2$ coefficients are zero at this level.

\begin{figure}[t]
\begin{center}
\subfigure[]{
\includegraphics[width=0.25\textwidth, bb= 0 0 338 209]{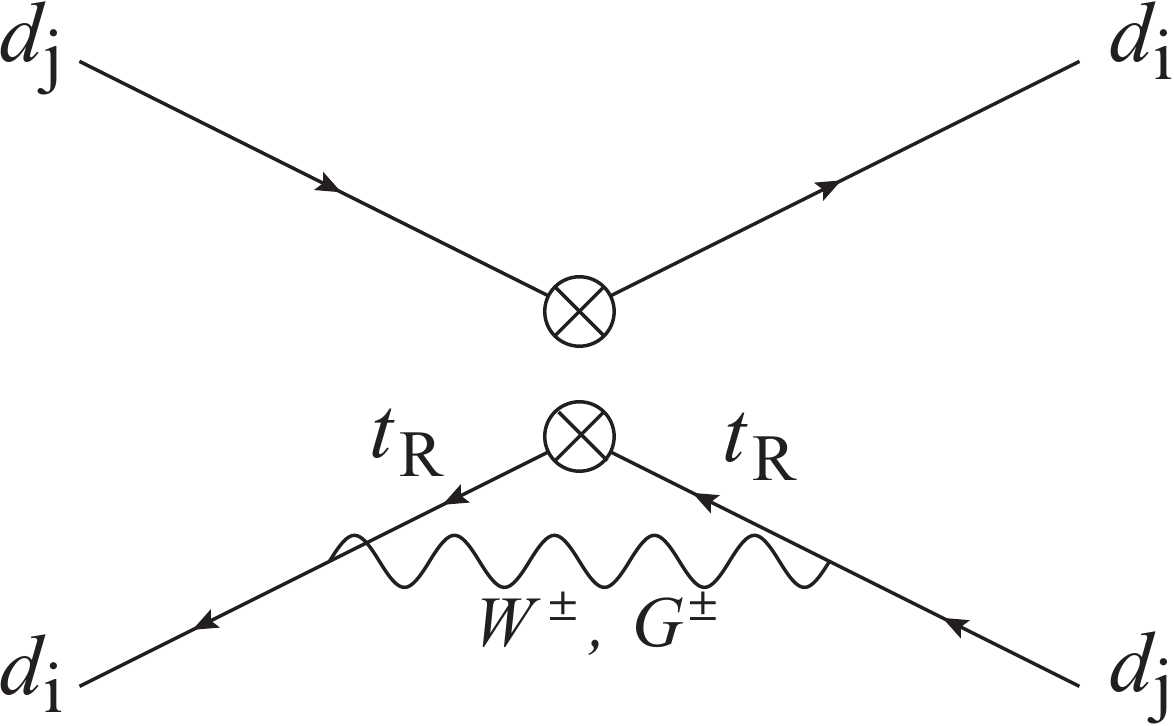}
}~
\subfigure[]{
\includegraphics[width=0.25\textwidth, bb= 0 0 338 208]{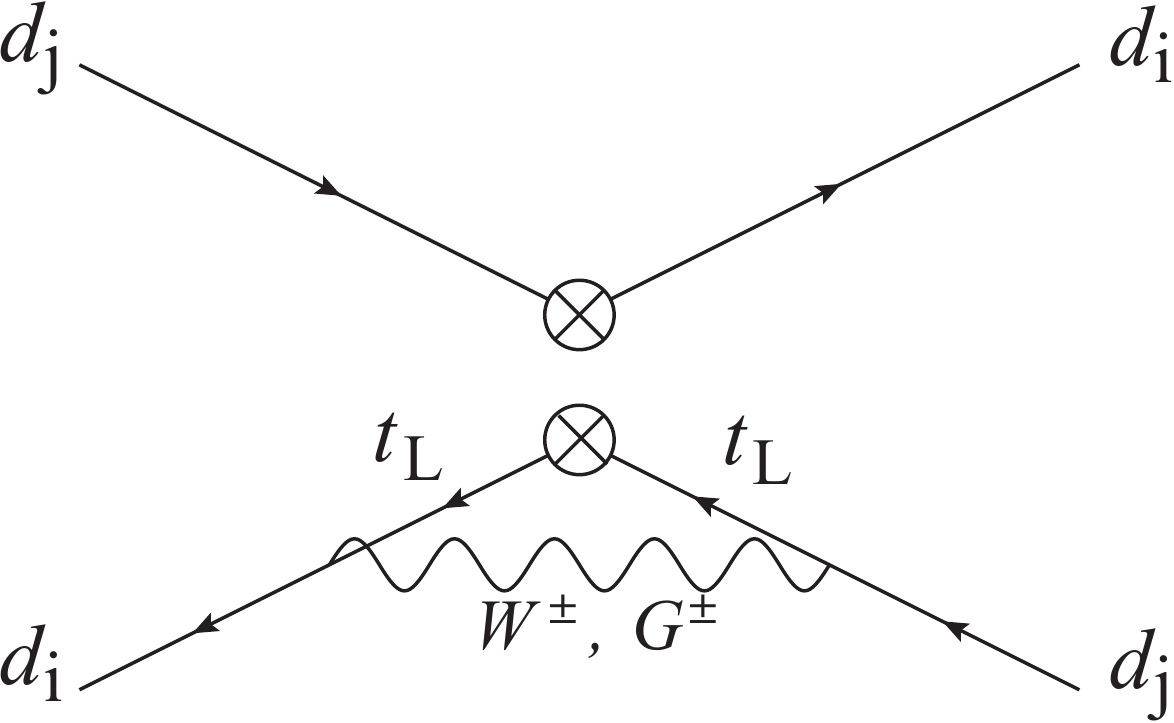}
}~
\subfigure[]{
\includegraphics[width=0.25\textwidth, bb= 0 0 338 211]{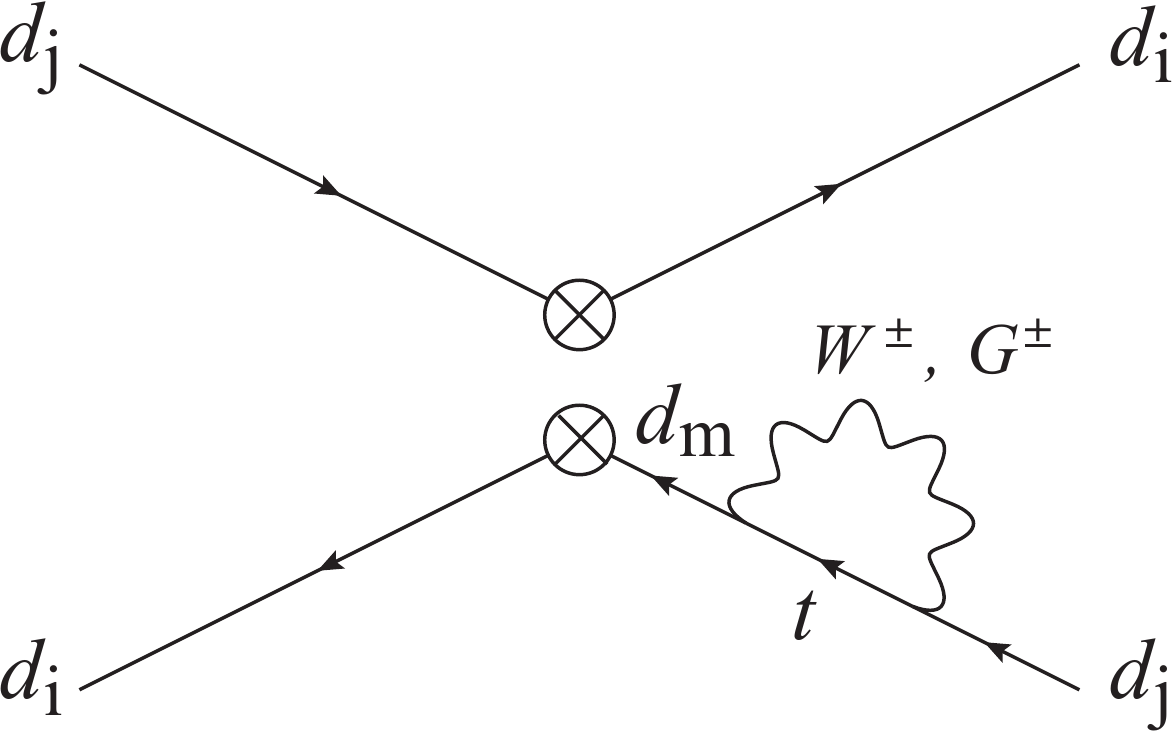}
}
\\
\vspace{.3 cm}
\subfigure[]{
\includegraphics[width=0.25\textwidth, bb= 0 0 338 209]{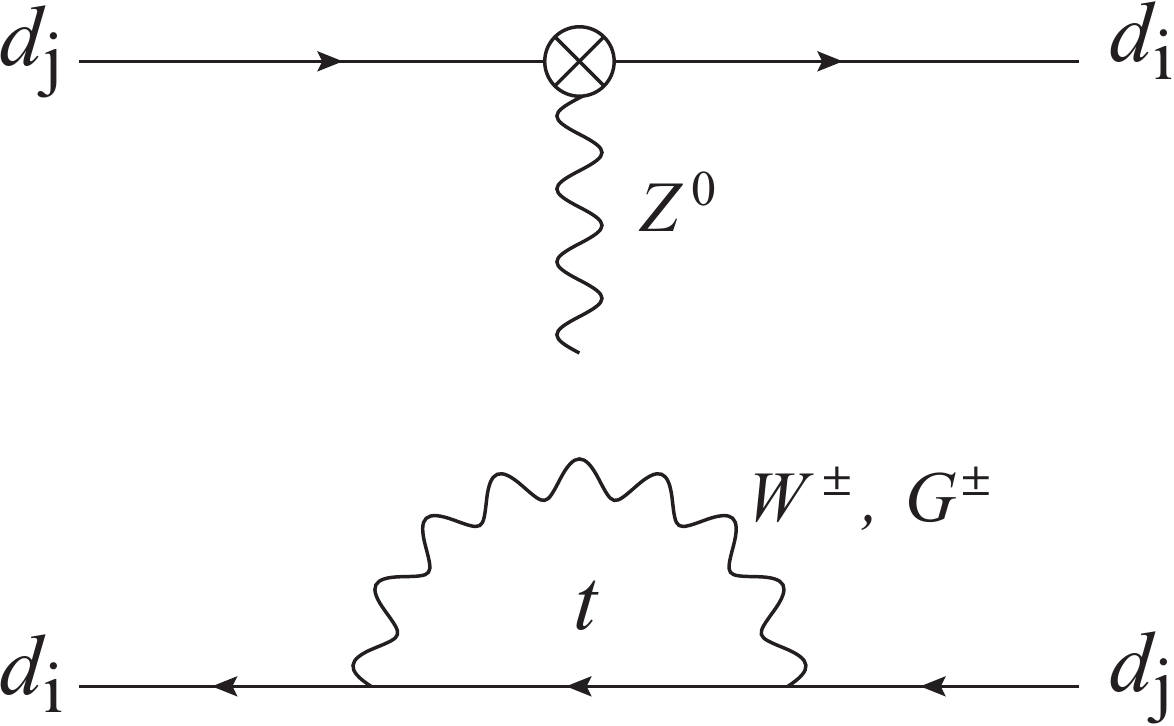}
}~
\subfigure[]{
\includegraphics[width=0.25\textwidth, bb= 0 0 338 209]{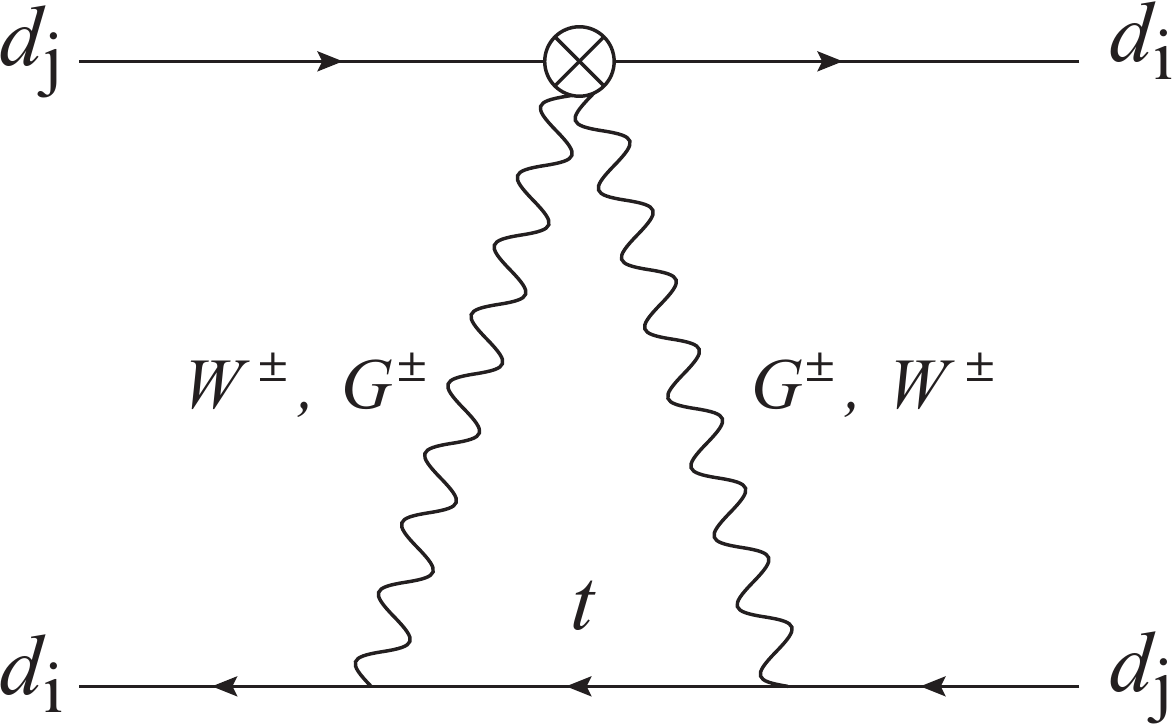}
}~
\subfigure[]{
\includegraphics[width=0.25\textwidth, bb= 0 0 338 210]{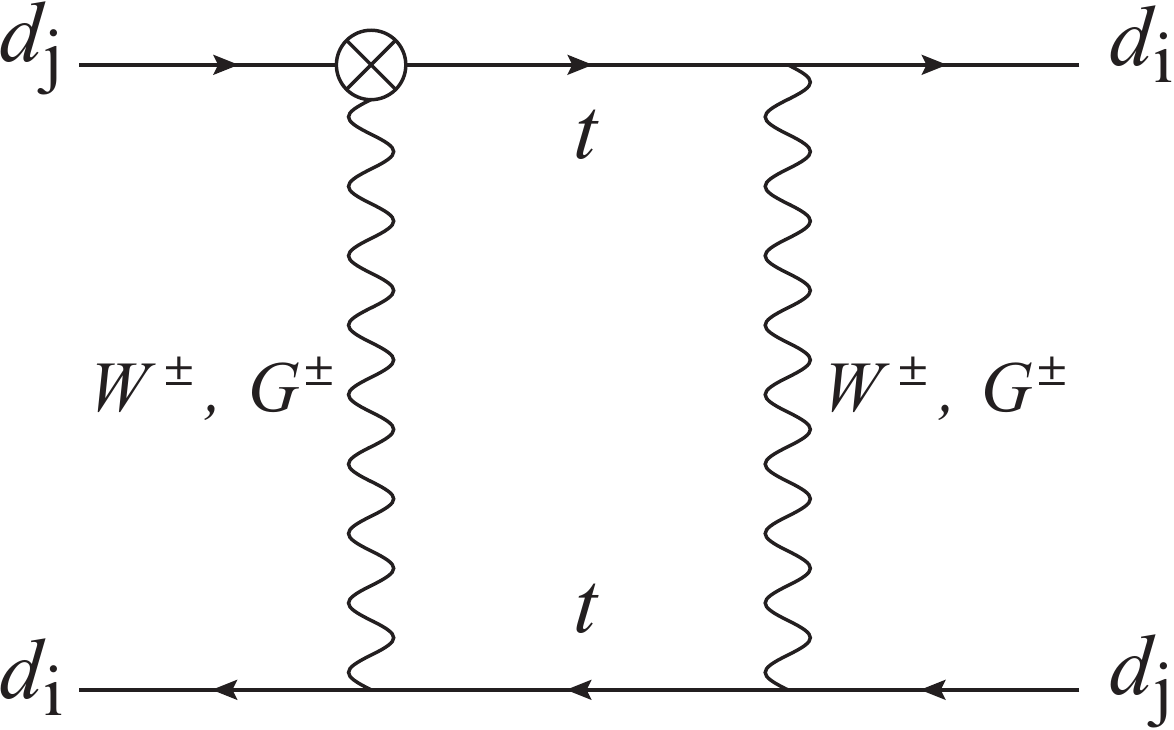}
}
\caption{
Feynman diagrams for the one-loop matchings onto the $\Delta F = 2$ operators $(i\neq j)$.
}
\label{fig:diagrams}
\end{center}
\end{figure}

Radiative corrections from the top quark can be sizable because of the large Yukawa coupling.
Combined with the SM bosons, they contribute to flavor-changing (FC) transitions of the down-type quarks. 
In particular, the SMEFT $\Delta F=1$ operators can induce the $\Delta F=2$ amplitudes through the RGEs and the one-loop matchings at the weak scale, which are exhibited in Fig.~\ref{fig:diagrams}.
The one-loop matching conditions in the Feynman-'t Hooft gauge are obtained as
\begin{align}
 (C_1)_{ij}^{\rm 1\textrm{--}loop} &=
 \frac{\alpha \lambda_t^{ij}}{4\pi s_W^2} \left[
 \left( -2+\frac{2}{N_c} \right) 
     (C^{(8)}_{qu})_{ij33} 
 - 4 (C^{(1)}_{qu})_{ij33} 
 + 4 (C^{(1)}_{Hq})_{ij} \right] I_1(x_t,\mu_{W})
 \notag \\ & 
 - \frac{2\alpha \lambda_t^{ij}}{\pi s_W^2} \bigg[
   (C^{(1)}_{qq})_{ij33} 
 + (C^{(1)}_{qq})_{33ij} 
 - (C^{(3)}_{qq})_{ij33} 
 - (C^{(3)}_{qq})_{33ij}
 + 2 (C^{(3)}_{qq})_{3ji3} 
 + 2 (C^{(3)}_{qq})_{i33j} \bigg] J(x_t)
 \notag \\ & 
 + \frac{\alpha}{2\pi s_W^2} \sum_{m=1}^{3}  \bigg[
   \lambda_t^{im} \Big(
   (C^{(1)}_{qq})_{mjij} 
 + (C^{(1)}_{qq})_{ijmj} 
 + (C^{(3)}_{qq})_{mjij} 
 + (C^{(3)}_{qq})_{ijmj} \Big)
 \notag \\ & \quad\quad\quad\quad\quad
 + \lambda_t^{mj} \Big(
   (C^{(1)}_{qq})_{imij} 
 + (C^{(1)}_{qq})_{ijim}
 + (C^{(3)}_{qq})_{imij} 
 + (C^{(3)}_{qq})_{ijim} \Big) \bigg] K(x_t,\mu_{W})
 \notag \\ & 
 - \frac{\alpha \lambda_t^{ij}}{\pi s_W^2} (C^{(3)}_{Hq})_{ij} I_2(x_t,\mu_{W})
 + \frac{\alpha \lambda_t^{ij}}{4 \pi s_W^2} \sum_{m=1}^{3} 
 \left[ 
    \lambda_t^{im} (C^{(3)}_{Hq})_{mj}
  + (C^{(3)}_{Hq})_{im} \lambda_t^{mj}
 \right] S_0(x_t),
 \label{eq:EWSBmatchingLoop1} \\
 (C_4)_{ij}^{\rm 1\textrm{--}loop} &=
   \frac{  \alpha \lambda_t^{ij}}{\pi s_W^2} (C^{(8)}_{ud})_{33ij} I_1(x_t,\mu_{W})
 + \frac{2 \alpha \lambda_t^{ij}}{\pi s_W^2} (C^{(8)}_{qd})_{33ij} J(x_t)
 \notag \\ & 
 - \frac{\alpha}{2\pi s_W^2} \sum_{m=1}^{3} \bigg[
   \lambda_t^{im} (C^{(8)}_{qd})_{mjij} 
 + \lambda_t^{mj} (C^{(8)}_{qd})_{imij} \bigg] K(x_t,\mu_{W}),
 \\
 (C_5)_{ij}^{\rm 1\textrm{--}loop} &=
 \frac{2 \alpha \lambda_t^{ij}}{\pi s_W^2} \left[
                  (C^{(1)}_{ud})_{33ij} 
 - \frac{1}{2N_c} (C^{(8)}_{ud})_{33ij} 
 -                (C_{Hd})_{ij} \right] I_1(x_t,\mu_{W})
 \notag \\ &
 + \frac{4 \alpha \lambda_t^{ij}}{\pi s_W^2} \left[
                  (C^{(1)}_{qd})_{33ij}
 - \frac{1}{2N_c} (C^{(8)}_{qd})_{33ij} \right] J(x_t)
 \notag \\ & 
 - \frac{\alpha}{\pi s_W^2} \sum_{m=1}^{3} \bigg[
   \lambda_t^{im} \Big(
                  (C^{(1)}_{qd})_{mjij}
 - \frac{1}{2N_c} (C^{(8)}_{qd})_{mjij} \Big) 
 \notag \\ & ~~~~~~~~~~~~~~~
 + \lambda_t^{mj} \Big( 
                  (C^{(1)}_{qd})_{imij}
 - \frac{1}{2N_c} (C^{(8)}_{qd})_{imij} \Big) \bigg]
 K(x_t,\mu_{W}),
 \label{eq:EWSBmatchingLoop3}
\end{align}
where the parameters are defined as
\begin{align}
 x_t &\equiv \frac{m_t^2}{M_W^2},~~~
 \lambda_t^{ij} \equiv V_{ti}^{\ast} V_{tj}.
\end{align}
Here, $V_{ij}$ is the CKM matrix, and $s_W = \sin\theta_W$ with the Weinberg angle $\theta_W$.
The loop functions are given as
\begin{align}
 I_1(x,\mu) &=
 \frac{x}{8} \left[ 
 \ln\frac{\mu}{M_W} - \frac{x-7}{4(x-1)} - \frac{x^2-2x+4}{2(x-1)^2} \ln x 
 \right],
 \\
 I_2(x,\mu) &= 
 \frac{x}{8} \left[
 \ln \frac{\mu}{M_W} + \frac{7 x -25}{4(x-1)} - \frac{x^2-14x+4}{2(x-1)^2} \ln x 
 \right],
 \\
 J(x) &= 
 \frac{x}{16} \left( 1 - \frac{2\ln x}{x-1} \right),
 \\
 K(x,\mu) &= 
 \frac{x}{8} \left[
 \ln \frac{\mu}{M_W} + \frac{3(x+1)}{4(x-1)} - \frac{x(x+2)}{2(x-1)^2} \ln x
 \right],
 \\
 S_0(x) &=
 \frac{x}{4} \left[ 
 \frac{x^2-11x+4}{(x-1)^2} + \frac{6x^2}{(x-1)^3} \ln x
 \right].
\end{align}
Here, the $\overline{\text{MS}}$ regularization scheme is adopted.
In the result, the Wilson coefficients in the left-handed side are in the low-scale basis, and those in the right-handed side are in the SMEFT. 
Both of them are evaluated at the weak scale, $\mu=\mu_{W}$.
The other low-scale $\Delta F=2$ operators do not receive one-loop corrections through the top quark decoupling. 

The contributions from $\mathcal{O}^{(1,8)}_{qu}$ and $\mathcal{O}^{(1,8)}_{ud}$, in which the $W$ and NG bosons that couple to virtual top quarks are exchanged, are shown in Fig.~\ref{fig:diagrams}(a) and give a loop function $I_1(x,\mu)$.
Those from $\mathcal{O}^{(1,3)}_{qq}$ and $\mathcal{O}^{(1,8)}_{qd}$ are shown in Fig.~\ref{fig:diagrams}(b) and give $J(x)$.
Those with the $K(x,\mu)$ function come from FC self-energy corrections to the down-type quarks in the effective operators [Fig.~\ref{fig:diagrams}(c)], where the top quark is exchanged.
The results for the quark--Higgs operators, $\mathcal{O}^{(1,3)}_{Hq}$ and $\mathcal{O}_{Hd}$ [Fig.~\ref{fig:diagrams}(d--f)], are consistent with those in Refs.~\cite{Endo:2016tnu,Bobeth:2017xry} and give loop functions $I_1(x,\mu)$, $I_2(x,\mu)$ and $S_0 (x)$.
The loop functions, $I_1(x,\mu), I_2(x,\mu)$ and $K(x,\mu)$, depend on the matching scale $\mu$ explicitly. However, $J(x)$ is independent of $\mu$ at the $\mathcal{O}(y_t^2)$ level:
the scale dependence stems from Fig.~\ref{fig:diagrams}(b) at the $\mathcal{O}(g^2)$ level, which is discarded in our approximation.\footnote{
 Such a divergence is canceled in the SM due to the GIM mechanism. 
 In Fig.~\ref{fig:diagrams}(b), the GIM mechanism does not work because $\mathcal{O}^{(1,3)}_{qq}$ and $\mathcal{O}^{(1,8)}_{qd}$ depend on the up-type quark flavor. 
} 
We checked that this logarithmic dependence is consistent with the anomalous dimensions in Ref.~\cite{Jenkins:2013zja,Jenkins:2013wua,Alonso:2013hga}.
As a result, the logarithmic dependence on $\mu_{W}$ cancels out by taking account of the RGEs in the leading-logarithmic accuracy.\footnote{
Focusing on the top-Yukawa terms,  we checked the following relations  in the leading-logarithmic accuracy,
\beq
\frac{\partial (C_{1,4,5})_{ij}}{\partial \ln \mu_{W}} =  \frac{\partial (C_{1,4,5})_{ij}^{\textrm{tree}}}{\partial \ln \mu_{W}} + \frac{\partial (C_{1,4,5})_{ij}^{\textrm{1-loop}}}{\partial \ln \mu_{W}} = 0.
\eeq
}\footnote{
The regularization scheme dependences generally exist in the one-loop matching conditions. 
 They must be canceled by including the SMEFT two-loop RGEs, which is beyond the scope of this paper. 
 See Ref.~\cite{Bobeth:2017xry} for a discussion on the scheme dependence.
}
This is expected because this dependence in the matching conditions has the same origin as the beta functions in calculating loop diagrams (see Ref.~\cite{Endo:2016tnu}).

The double-penguin contributions should vanish when the gauge bosons of the SM unbroken gauge symmetries, $g$ and $\gamma$, are exchanged. This is guaranteed by the unbroken gauge symmetry.
In Appendix~\ref{sec:DoublePenguin}, the cancellations are shown explicitly by using the one-loop matching conditions, Eqs.~\eqref{eq:EWSBmatchingLoop1}--\eqref{eq:EWSBmatchingLoop3}.
This justifies our result.

We adopted the Feynman-'t Hooft gauge for evaluating the weak gauge bosons. 
The low-scale Wilson coefficients should be gauge invariant under the SM weak gauge symmetry, because those gauge bosons are decoupled below the EWSB scale. 
In fact, the gauge invariance has been proven explicitly for the contributions of the quark--Higgs operators $(\mathcal{O}^{(1,3)}_{Hq})_{ij}$ and $(\mathcal{O}_{Hd})_{ij}$ by using the $R_\xi$ gauge in Ref.~\cite{Bobeth:2017xry}, which are shown with the loop functions $I_1(x,\mu)$, $I_2(x,\mu)$ and $S_0 (x)$ in Eqs.~\eqref{eq:EWSBmatchingLoop1}--\eqref{eq:EWSBmatchingLoop3}.

After matching onto the low-scale operators, they are evolved by the  RGEs as usual.
Then, the results are compared with the experimental data, i.e., the $K^0\textrm{--}\overline{K}{}^0$ and $B^0_q\textrm{--}\overline{B}{}^0_q$ $(q=d,s)$ oscillations.\footnote{
 Our analysis is performed for the $\Delta F=2$ processes of the down-type quarks.
 The SMEFT $\Delta F=1$ operators can also contribute to those of the up-type quarks, e.g., the $D^0\textrm{--}\overline{D}{}^0$ oscillation. 
 Then, the bottom quark loops appear in Fig.~\ref{fig:diagrams}, and the one-loop corrections would be proportional to the bottom Yukawa coupling instead of the top one.
}

\section{Left-right symmetric models}

In this section, let us study left-right symmetric models to demonstrate the SMEFT corrections of the dynamical top quark as explored in Sec.~\ref{sec:formula}. 
In particular, we focus on the effects of the SMEFT $\Delta F=1$ operators for the $\Delta F=2$ transitions.
 
The left-right extension of the SM implements the parity violation in the weak interaction by spontaneously breaking the ${\rm SU(3)}_C \times {\rm SU(2)}_L \times {\rm SU(2)}_R \times {\rm U(1)}_{B-L}$ gauge symmetries~\cite{Pati:1974yy,Mohapatra:1974hk,Mohapatra:1974gc,Senjanovic:1975rk,Senjanovic:1978ev}.
The new right-handed $W$ boson generates FC charged currents in addition to the SM left-handed $W$ boson. 
The quark interactions of the left- and right-handed $W$ bosons are
\begin{align}
 \mathcal{L}_{\rm int} = 
   \frac{g_L}{\sqrt{2}} (V_{L})_{ij} \bar{u}_{i} \gamma_{\mu} P_L d_{j} W_{L}^{\mu}
 + \frac{g_R}{\sqrt{2}} (V_{R})_{ij} \bar{u}_{i} \gamma_{\mu} P_R d_{j} W_{R}^{\mu}
 + {\rm h.c.},
\end{align}
where the first term is for the SM $W$ boson.
The right-handed $W$ boson, $W_R$, is obtained by replacing $L \leftrightarrow R$, in the second term.
Here, the new coupling $g_R$ and the mixing matrix $V_{R}$ are introduced for $W_R$ similarly to $W_L$. 

The gauge symmetries are broken to ${\rm SU(3)}_C \times {\rm U(1)}_{em}$ by Higgs vacuum expectation values (VEVs). 
In the minimal setup, the VEV of the Higgs field, $\Delta_R$, whose charges are $({\rm SU(2)}_L, {\rm SU(2)}_R, {\rm U(1)}_{B-L}) = (1,3,2)$, breaks the left-right symmetry, ${\rm SU(2)}_L \times {\rm SU(2)}_R \times {\rm U(1)}_{B-L}$, to ${\rm SU(2)}_L \times {\rm U(1)}_{Y}$.
The VEV of the Higgs bi-doublet, $\Phi\in(2,2,0)$, enables EWSB. 
On the other hand, the VEV of $\Delta_L \in (3,1,2)$ is assumed to be suppressed.
Their components are expressed as
\begin{align}
 \Delta_i = 
 \begin{bmatrix}
  \Delta_i^+/\sqrt{2} & \Delta_i^{++} \\
  \Delta_i^0 & -\Delta_i^+/\sqrt{2} 
 \end{bmatrix}
 ~(i=L,R),~~~
 \Phi = 
 \begin{bmatrix}
  \phi_1^0 & \phi_2^+ \\
  \phi_1^- & \phi_2^0 
 \end{bmatrix}.
\end{align}
The spontaneous symmetry breaking is achieved by the VEVs,
\begin{align}
 \langle\Delta_{L,R}\rangle = 
 \frac{1}{\sqrt{2}} 
 \begin{bmatrix}
  0 & 0 \\
  v_{L,R} & 0 
 \end{bmatrix},~~~
 \langle\Phi\rangle = 
 \frac{1}{\sqrt{2}}
 \begin{bmatrix}
  v \cos\beta & 0 \\
  0 & v \sin\beta\, e^{i\alpha} 
 \end{bmatrix}.
 \label{eq:higgsVEV}
\end{align}
We impose a hierarchy among the Higgs VEVs, $v_R \gg v \cos \beta, v \sin \beta \gg v_L$, in order to be consistent with observed phenomena and to avoid fine-tunings in the scalar potential \cite{Gunion:1989in, Deshpande:1990ip}.
An angle $\alpha$ is a spontaneous $CP$-violating phase.
In addition to the QCD $\theta$ term, $\alpha$ induces the strong $CP$ phase~\cite{Maiezza:2014ala}\footnote{
 See discussions in Refs.~\cite{Mohapatra:1978fy,Babu:1989rb,Barr:1991qx} for the strong $CP$ problem with a generalized parity invariance $\mathcal{P}$.
}, which is severely constrained by the neutron electric dipole moment \cite{Afach:2015sja}.
As we will see below, the following analysis is independent of $\alpha$. 
The masses of the left and right-handed $W$ bosons are approximately given by
\begin{align}
 M_{W_L}^2 \simeq \frac{g_L^2}{4}v^2,~~~
 M_{W_R}^2 \simeq \frac{g_R^2}{2}v_R^2,
\end{align}
for $v_R \gg v$ with $v \simeq 246\,{\rm GeV}$.

In addition to the $W$ bosons, heavy Higgs bosons, $H^0$ and $H^{\pm}$, have FC couplings as 
\begin{align}
 -\mathcal{L}_{\rm int} &\simeq
 \frac{\sqrt{2}}{v \cos 2\beta} 
 \Big[ 
   \bar{d} (V_L^{\dagger} M_u V_R) P_R d\, H^0  
 + \bar{d} (V_R^{\dagger} M_u V_L) P_L d\, (H^0)^{\ast}
 \notag \\ & \quad\quad\quad\quad\quad
 + \bar{u} (M_u V_R)           P_R d\, H^{+} 
 + \bar{d} (V_R^{\dagger} M_u) P_L u\, H^{-} \Big],
 \label{eq:FCH}
\end{align}
with $H^{0} = \cos \beta \phi_2^0 - \sin \beta  e^{ i \alpha}\left( \phi_1^{0}\right)^{\ast}$ and 
$H^+ = \cos \beta \phi_2^+ + \sin \beta e^{ i \alpha} \phi_1^{+}$. 
Here, $v_R \gg v$ is assumed, and the up-type quark masses is $M_u = {\rm diag}(m_u, m_c, m_t)$.
The masses of the heavy Higgs bosons, $M_H$, are almost proportional to $v_R$.
The Higgs potential in the limit of $v_R \gg v$ is given in Appendix~\ref{sec:Higgssector}.

The right-handed $W$ boson and the heavy neutral Higgs boson, as well as the SM (left-handed) $W$ boson, induce $\Delta F=2$ transitions~\cite{Beall:1981ze,Mohapatra:1983ae}.
They are severely constrained by the observed meson oscillations.
First of all, let us briefly overview the conventional approach.
In literature, the Wilson coefficients of the low-scale operators in Eq.~\eqref{Hamiltonianeff} are set by integrating out $W_R$ and $H^0$ as well as $W_L$ and the up-type quarks~\cite{Basecq:1985cr,Bertolini:2014sua}:
\begin{align}
 (C_{4})_{ij}^{H\textrm{--}{\rm tree}} &= 
 -\frac{2\sqrt{2} G_F}{\cos^2 2\beta} \sum_{k,l}
 \frac{m_{u_k} m_{u_l}}{M_H^2} (\lambda^{LR})_k^{ij} (\lambda^{RL})_l^{ij},
 \label{eq:H_tree}
 \\
 (C_{4})_{ij}^{W_L\textrm{--}W_R} &= 
 \frac{g_L^2 g_R^2}{16 \pi^2} \sum_{k,l}
 \frac{m_{u_k} m_{u_l}}{M_{W_L}^2 M_{W_R}^2} (\lambda^{LR})_k^{ij} (\lambda^{RL})_l^{ij}\,
 \mathcal{F}_A (x_k,x_l,\beta),
 \\
 (C_{4})_{ij}^{H\textrm{--}{\rm s.e.}} &= 
 -\frac{g_L^2 g_R^2}{128 \pi^2} \sum_{k,l}
 \frac{m_{u_k} m_{u_l}}{M_{W_L}^2 M_{W_R}^2} (\lambda^{LR})_k^{ij} (\lambda^{RL})_l^{ij}\,
 \mathcal{F}_B (\tau_L,\tau_R),
  \label{eq:H_se}
 \\
 (C_{4})_{ij}^{H\textrm{--}{\rm vert.}} &= 
 -\frac{g_L^2 g_R^2}{16\pi^2} \sum_{k,l}
 \frac{m_{u_k} m_{u_l}}{M_{W_L}^2 M_{W_R}^2} (\lambda^{LR})_k^{ij} (\lambda^{RL})_l^{ij}\,
 \mathcal{F}_C (\tau_k,\tau_l,\tau_L,\tau_R),
 \label{eq:H_vert}
\end{align}
where the parameters are defined as
\begin{align}
 & 
 (\lambda^{LR})_k^{ij} \equiv (V_L^*)_{ki} (V_R)_{kj},~
 x_{k} \equiv \frac{m_{u_{k}}^2}{M_{W_L}^2},
 \notag \\ &
 \beta \equiv \frac{M_{W_L}^2}{M_{W_R}^2},~
 \tau_L \equiv \frac{M_{W_L}^2}{M_H^2},~
 \tau_R \equiv \frac{M_{W_R}^2}{M_H^2},~
 \tau_k \equiv \frac{m_{u_k}^2}{M_H^2},
\end{align}
and $(\lambda^{RL})_k^{ij}$ is given by replacing $L \leftrightarrow R$ in $(\lambda^{LR})_k^{ij}$. 
Here, the indices $k,l$ are the up-type quark flavor, and the definitions of the loop functions $\mathcal{F}_A$, $\mathcal{F}_B$ and $\mathcal{F}_C$ are summarized in Appendix~\ref{sec:loop}.\footnote{
 Our results in  Eqs.~\eqref{eq:H_se} and \eqref{eq:H_vert} are smaller than the result of Ref.~\cite{Bertolini:2014sua} by a factor of $2$.
}

Among the Wilson coefficients, the tree-level contribution, $(C_{4})^{H\textrm{--}{\rm tree}}$, is obtained by exchanging the heavy neutral Higgs boson. 
The one-loop contributions, $(C_{4})^{H\textrm{--}{\rm s.e.}}$ and $(C_{4})^{H\textrm{--}{\rm vert.}}$, are given by self-energy (s.e.) and vertex (vert.) corrections to the tree-level heavy neutral Higgs diagram, respectively. 
Here, the on-shell renormalization scheme is applied~\cite{Basecq:1985cr}.
On the other hand, the one-loop contribution $(C_{4})^{W_L\textrm{--}W_R}$ comes from a box diagram where both the left- and right-handed $W$ bosons as well as the up-type quarks are exchanged.\footnote{
 If $W_R$ and $H$ are sufficiently heavier than $W_L$, the $W_R$--$W_R$ box contribution is much smaller than the $W_L$--$W_R$ box one. 
} 
It is impoartant that $(C_{4})^{W_L\textrm{--}W_R}$ itself depends on a choice of the gauge fixing. 
Here and hereafter, the Feynman-'t Hooft gauge is used. 
The gauge invariance of the transition amplitude is guaranteed by adding the one-loop neutral Higgs contributions, $(C_{4})^{H\textrm{--}{\rm s.e.}}$ and $(C_{4})^{H\textrm{--}{\rm vert.}}$~\cite{Chang:1984hr,Basecq:1985cr,Hou:1985ur,Ecker:1985vv}.

In the conventional calculation (Ref.~\cite{Bertolini:2014sua} as a representative case), after the above Wilson coefficients are set, the RGEs for the low-scale operators are solved \cite{Buras:2001ra}.
However, it is noticed that the one-loop diagrams include the left-handed $W$ boson and the up-type quarks, which are much lighter than the right-handed $W$ and heavy Higgs bosons for, e.g., the LHC constraints~\cite{LHC,Maiezza:2010ic}. 
Hence, it is uncertain in which energy scale the Wilson coefficients should be input.
Moreover, the heavy charged Higgs boson contributes to the $\Delta F=2$ transitions through box diagrams with the SM W boson and the up-type quarks. 
Although the contribution is often neglected in the literature (see Ref.~\cite{Ecker:1985vv} for an early work), it may be comparable to $(C_{4})^{H\textrm{--}{\rm s.e.}}$ and $(C_{4})^{H\textrm{--}{\rm vert.}}$.
Since the SM W boson and the up-type quarks are much lighter than the heavy charged Higgs boson, the scale uncertainty problem arises similarly to the above.
In the following, we study the $\Delta F=2$ processes in left-right symmetric models by the procedure explored in Sec.~\ref{sec:formula}.
\\

\begin{figure}[t]
\begin{center}
\includegraphics[width=.9\textwidth, bb= 0 0 999 658]{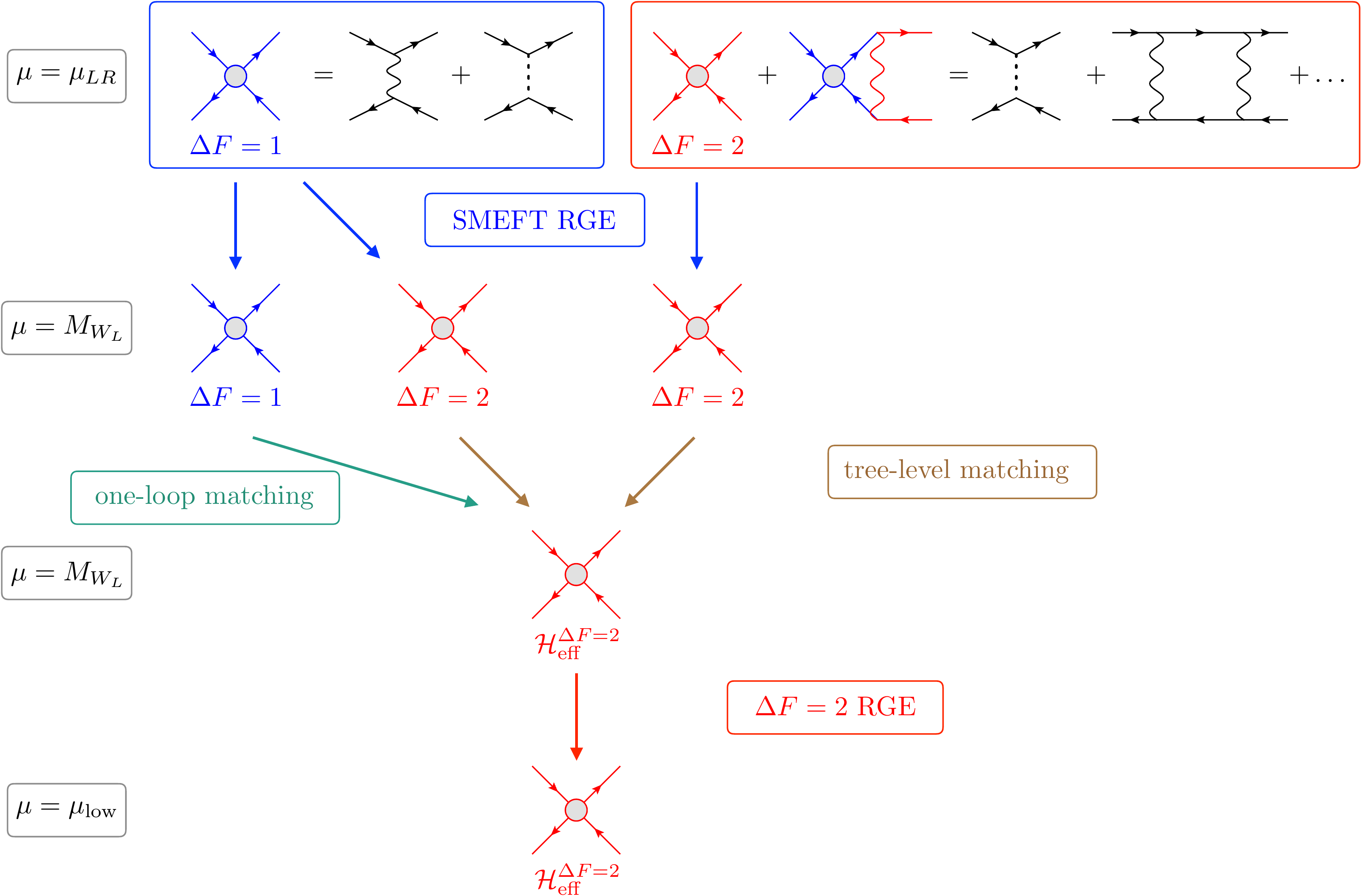}
\vspace{0.1cm}
\caption{
Schematic figure for the SMEFT framework in the left-right symmetric model.
}
\label{fig:framework}
\end{center}
\end{figure}

In this paper, we focus on the top quark contribution as mentioned in Sec.~\ref{sec:introduction}.
First of all, let us summarize the analysis procedure in Fig.~\ref{fig:framework}. 
At the decoupling scale of the left-right symmetry $(\mu_{LR})$, the Wilson coefficients in the SMEFT are evaluated.
In addition to the $\Delta F =2$ operators (the red colored diagrams in Fig.~\ref{fig:framework}), there are $\Delta F=1$ top-quark operators which eventually contribute to the $\Delta F =2$ transitions (the blue colored diagrams).
After solving the SMEFT RGEs, they are matched onto the low-scale operators at the EWSB scale, where we need to take account of the one-loop level matching condition.\footnote{
The SMEFT $\Delta F =2$ operators are matched onto the low-scale $\Delta F =2$ weak Hamiltonaian at the tree level, because there are no $\mathcal{O}(y_t^2)$ contributions to the one-loop matching conditions for them.
 On the other hand, in order to investigate an impact of the top Yukawa contributions, 
 we discarded the $\mathcal{O}(\alpha_s)$ one-loop matching contributions, which does not change the flavor.
}
Below the EWSB scale, we follow the standard procedure for the $\Delta F =2$ observables.

First, let us consider the matching condition of the SMEFT at $\mu=\mu_{LR}$ (the first line in the Fig.~\ref{fig:framework}).
At the tree level, one obtains the following $\Delta F=1$ SMEFT operators at the dimension six after integrating out $W_R$,  
\begin{align}
 (C^{(8)}_{ud})_{33ij}^{\rm tree} &= - 
 \frac{g_R^2}{M_{W_R}^2}(V_R^{\dagger})_{i3} (V_R)_{3j},
 \\
 (C^{(1)}_{ud})_{33ij}^{\rm tree} &= \frac{1}{2N_c} (C^{(8)}_{ud})_{33ij}.
\end{align}
In addition, by exchanging the heavy neutral and charged Higgs bosons, we obtain the following $\Delta F=1$ operators,
\begin{align}
 (C_{qd}^{(8)})_{3 3 i j}^{\textrm{tree}} &= 
 - \frac{2\sqrt{2} G_F}{\cos^2 2\beta} 
 \frac{m_t^2}{M^2_{H}}
 (V_R^{\dag})_{i 3} (V_R)_{3j},
 \label{eq:chH1}\\
 (C_{qd}^{(1)})_{3 3 i j}^{\textrm{tree}} &= 
 \frac{1}{2 N_c} (C_{qd}^{(8)})_{33ij}.
 \label{eq:chH2}
\end{align}
The details of the calculations are found in Appendix~\ref{sec:Higgssector}. 
On the other hand, the $\Delta F =2$ SMEFT operators are derived at the tree level from the exchange of the heavy neutral Higgs bosons as
\begin{align}
 (C^{(8)}_{qd})_{ijij}^{\rm tree} &= 
 - \frac{2\sqrt{2} G_F}{\cos^2 2\beta} \frac{m_t^2}{M_H^2} 
 (\lambda^{LR})_t^{ij} (\lambda^{RL})_t^{ij},
 \\
 (C^{(1)}_{qd})_{ijij}^{\rm tree} &= \frac{1}{2N_c} (C^{(8)}_{qd})_{ijij}. 
\end{align}
All the above tree-level Wilson coefficients are evaluated at $\mu = \mu_{LR}$.

As for the one-loop level matching, the self-energy and vertex corrections of the heavy neutral Higgs discussed above contribute to the $\Delta F = 2$ Wilson coefficients. 
Besides, in discussing the $W_L$--$W_R$ box contributions, one needs to avoid double counting from  the one-loop contribution with $(C^{(8)}_{ud})_{33ij}$, where the top-quark loop is enclosed by the SM $W$ boson. 
The result is obtained as
\begin{align}
 (C^{(8)}_{qd})_{ijij}^{\rm 1\textrm{--}loop} =\,& 
 \frac{g_L^2 g_R^2 m_t^2}{16 \pi^2 M_{W_L}^2 M_{W_R}^2} 
 (\lambda^{LR})_t^{ij} (\lambda^{RL})_t^{ij}
 \bigg[
   \mathcal{F}_A (x_t,x_t,\beta)
 - \frac{1}{8} \mathcal{F}_B (\tau_L,\tau_R)
 - \mathcal{F}_C (\tau_t,\tau_t,\tau_L,\tau_R) 
 \bigg]
 \notag \\ &
 + \frac{g_L^2 g_R^2}{4\pi^2 M_{W_R}^2} 
 (\lambda^{LR})_t^{ij} (\lambda^{RL})_t^{ij} I_1(x_t,\mu_{LR}),
 \\
 (C^{(1)}_{qd})_{ijij}^{\rm 1\textrm{--}loop} =\,& 
 \frac{1}{2N_c} (C^{(8)}_{qd})_{ijij},
\end{align}
where the second term of $C^{(8)}_{qd}$ stands for the subtraction to avoid the double counting. 
We can see that the $\mu_{LR}$ dependence in $(C^{(8)}_{qd})_{ijij}^{\rm 1\textrm{--}loop}$ is dropped when the scale is set to be $\mu_{LR} = M_{W_R}$.
In addition, the one-loop matching condition that comes from the $H^{\pm}$ and $W_L$ box diagrams is obtained as
\begin{align}
 (C_{qd}^{(8)})_{ijij}^{\textrm{1-loop}} =\,&
 \frac{\sqrt{2} G_F}{\pi^2} \frac{g_L^2}{\cos^2 2\beta} \frac{m_t^2}{M_H^2}
 (\lambda^{LR})_{t}^{ij} (\lambda^{RL})_{t}^{ij} 
 \left[
 \frac{1}{16}\mathcal{F}_{D} \left( x_t, x_t, \tau_L \right)
 + J(x_t)
 \right],
 \label{eq:ChargedHiggsLoop} \\ 
 (C_{qd}^{(1)})_{ijij}^{\textrm{1-loop}} =\,& 
 \frac{1}{2 N_c} (C_{qd}^{(8)})_{ijij},
\end{align}
where the loop function $\mathcal{F}_{D}$ defined in Appendix~\ref{sec:loop} comes from $H^{\pm}$--$W_L$ box diagrams, whose result is consistent with that in Ref.~\cite{Ecker:1985vv}.
The contribution $J(x_t)$ is from the subtraction to avoid the double counting in similar to the $W_R$ case.
All the above Wilson coefficients for the one-loop level matching conditions are evaluated at $\mu = \mu_{LR}$.

After setting the Wilson coefficients for the dimension-six SMEFT operators at the scale $\mu = \mu_{LR}$, the SMEFT RGEs are solved to the EWSB scale, for which we choose $\mu = M_{W_L}$ (the second line in the Fig.~\ref{fig:framework}).
The one-loop level RGEs are summarized in Appendix~\ref{sec:RGE}.
At the EWSB scale, the SMEFT operators are matched onto the low-scale ones (the third line).
The tree-level and one-loop level matching conditions are found in Eqs.~\eqref{eq:EWSBmatchingTree1}--\eqref{eq:EWSBmatchingTree4} and Eqs.~\eqref{eq:EWSBmatchingLoop1}--\eqref{eq:EWSBmatchingLoop3}, respectively.
After the EWSB matching, the calculations are performed as usual, i.e., in the same way as the conventional approach.

\begin{figure}[t]
\begin{center}
\subfigure[]{
\includegraphics[width=0.45\textwidth, bb= 0 0 360 358]{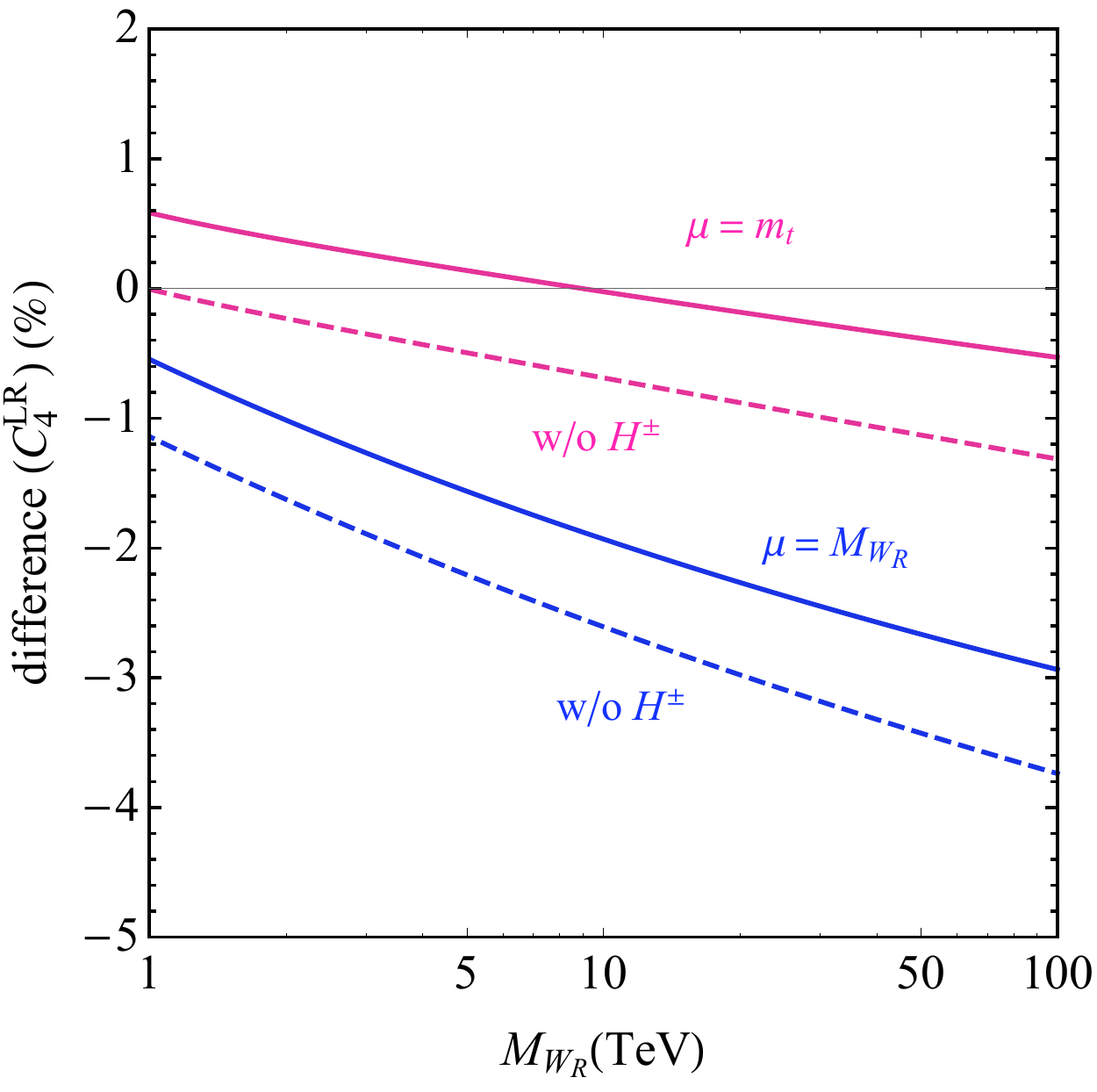}
}
\subfigure[]{
\includegraphics[width=0.45\textwidth, bb= 0 0 360 341]{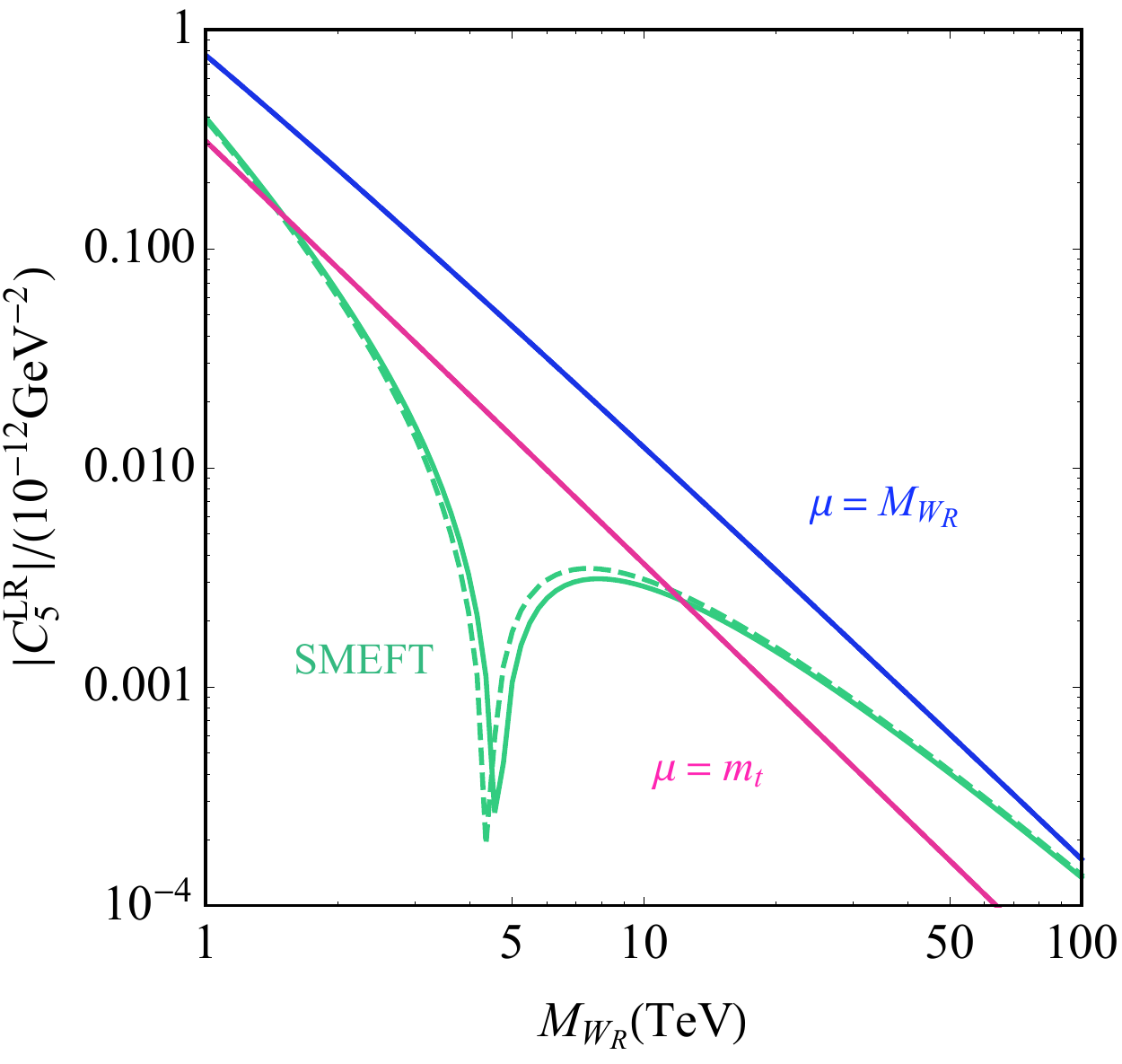}
}
\caption{
The low-scale Wilson coefficients $C_{4} (M_{W_L} )$ (left) and $C_{5} (M_{W_L} )$ (right) for $\Delta M_{B_s}$ in comparison with the conventional results.
In the conventional approach, the Wilson coefficients, \eqref{eq:H_tree}--\eqref{eq:H_vert}, are input at $\mu = M_{W_R}$ (blue) and $m_t$ (magenta).
The dashed lines do not include the contribution from the heavy charged Higgs boson.
}
\label{fig:DiffC45}
\end{center}
\end{figure}

\begin{figure}[t]
\begin{center}
\includegraphics[width=0.9\textwidth, bb= 0 0 447 296]{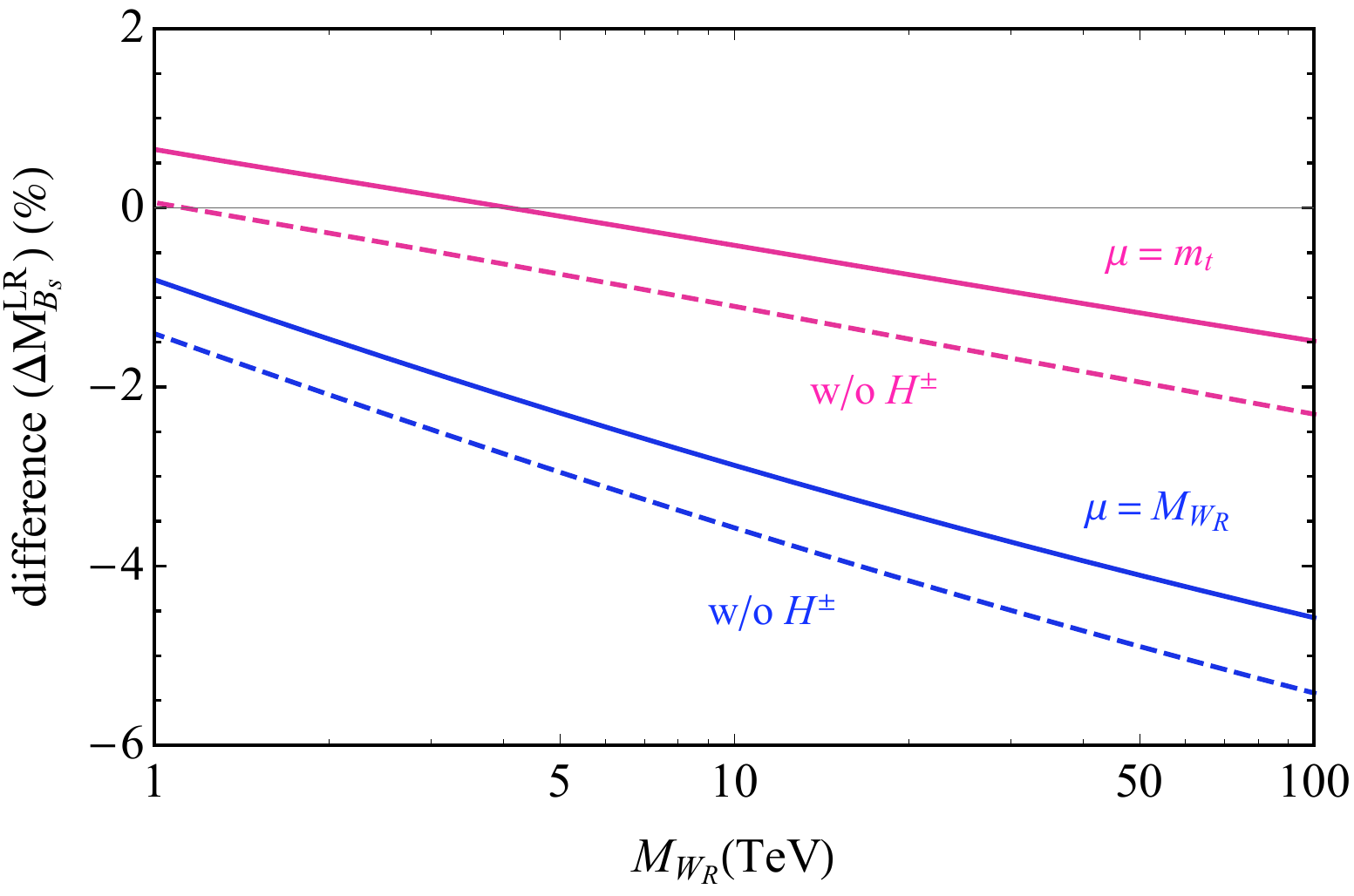}
\caption{
$\Delta M_{B_s}$ in comparison with the conventional results at $\mu = M_{W_R}$ (blue) and $m_t$ (magenta). The dashed lines do not include the contribution from the heavy charged Higgs boson.
}
\label{fig:DeltaMBs}
\end{center}
\end{figure}

The differences of our analysis from the conventional one are the SMEFT top-quark effects and the heavy charged Higgs boson contributions. 
In order to investigate their effects quantitatively, we consider the $\Delta B = 2$ process, $\Delta M_{B_s}$.
Let us define the difference as
\begin{align}
 \textrm{difference}\left(X \right) \equiv
 \frac{X^{\textrm{SMEFT}} - X^{\textrm{conventional}}}{X^{\textrm{SMEFT}}}
 \textrm{~~~~for~~~} 
 X = C_i(M_{W_L}),\, \Delta M_{B_s},
\end{align}
where $C_i(M_{W_L})$ is the low-scale Wilson coefficients at the EWSB scale for $\Delta M_{B_s}$, i.e., $i=3$ and $j=2$ in Eq.~\eqref{Hamiltonianeff}.
In the numerical analysis, we take $\tan \beta = m_b / m_t$, which naturally gives the fermion mass hierarchy $m_t \gg m_b$. 
The mass and scale are set as $M_H = 6 M_{W_R}$ and $\mu_{LR} = M_{W_{R}}$, respectively.
Also, we impose a generalized charge conjugation symmetry $\mathcal{C}$, which leads to $g_R = g_L$ and $V_R = K_u V_L^{\ast} K_d $ and $K_u = \textrm{diag}(e^{i \theta_{u}},e^{i \theta_{c}},e^{i \theta_{t}})$, $K_d = \textrm{diag}(e^{i \theta_{d}},e^{i \theta_{s}},e^{i \theta_{b}})$~\cite{Maiezza:2010ic, Maiezza:2014ala,Bertolini:2014sua}. 
In the evaluation of $\Delta M_{B_s}$, the latest lattice results \cite{Bazavov:2016nty} are applied for $B$-parameters.
We also use the {\tt RunDec} program \cite{Chetyrkin:2000yt} for evaluating the running top quark masses.

In Fig.~\ref{fig:DiffC45}(a), the difference of $C_{4} (M_{W_L})$ is shown.
The magenta and blue solid lines correspond to the cases of the conventional approach with different choices of the input scale of the Wilson coefficients.
Since it is uncertain in which energy scale the Wilson coefficients should be input, we set Eqs.~\eqref{eq:H_tree}--\eqref{eq:H_vert} at $\mu = M_{W_{R}}$ (blue) or at $\mu = m_t$ (magenta), and then, perform the low-scale RGEs to the lower scale. 
For instance, $\mu = m_t$ is chosen in Ref.~\cite{Bertolini:2014sua}.
It is found that the difference is less than three percents below $\mu_{LR} = 100\,{\rm TeV}$. 
Although $\mu = m_t$ seems to be favored for the conventional result, the deviation is enhanced as $\mu_{LR}$ increases. 

Our analysis includes both the top-quark effect and the heavy charged Higgs boson contribution. 
In order to investigate them individually, we show the results without introducing the latter contribution (dashed lines).
Hence, in Fig.~\ref{fig:DiffC45}(a), the deviations of the dashed lines from zero are due to the SMEFT top-quark effects explored in Sec.~\ref{sec:formula}.
It is found that the effects are less than four percents for $\mu_{LR} < 100\,{\rm TeV}$.
Also, the difference between the solid and dashed lines comes from the the heavy charged Higgs contribution. 
We confirm that it is about one percent level and is comparable to the one-loop contributions, $(C_{4})^{H\textrm{--}{\rm s.e.}}$ and $(C_{4})^{H\textrm{--}{\rm vert.}}$, in the Feynman-'t Hooft gauge.
The difference between the lines is insensitive to $M_{W_{R}}$, because the box contribution in Eq.~\eqref{eq:ChargedHiggsLoop}, i.e., the $\mathcal{F}_{D}$ term, dominates the total charged Higgs effects.
 
In Fig.~\ref{fig:DiffC45}(b), $C_{5} (M_{W_L})$ is displayed.
The magenta and blue solid lines correspond to $\mu = M_{W_{R}}$ and $\mu = m_t$ for the conventional approach, respectively.
In this case, $C_{5}$ is zero at the input scale and generated by $C_{4}$ through the RGEs down to $\mu = M_{W_L}$.
The dependence of $C_{5}$ on $M_{W_{R}}$ is thus from that of $C_{4}$.
The conventional analyses are compared with our SMEFT and $H^{\pm}$ results (green). 
The difference between the solid and dashed lines comes from the heavy charged Higgs boson, which is shown to be sub-leading similarly to the above case of $C_{4}$.
We found that $C_{5} (M_{W_L})$ depends heavily on $M_{W_{R}}$ and can be deviated from the conventional results by hundred percents.

In Fig.~\ref{fig:DeltaMBs}, the difference of $\Delta M_{B_s}$ is shown. 
Since it is dominated by $C_{4}$ at lower scales quantitatively, the result becomes similar to the one in Fig.~\ref{fig:DiffC45}(a).
It is seen that the SMEFT and charged Higgs effects are less than five percents for $\mu_{LR} < 100\,{\rm TeV}$ and are enhanced in larger $\mu_{LR}$. 
We also checked that these results are unchanged by a choice of $\theta_q$.
Also, we can derive the same conclusions for $\Delta M_{B_d}$ as $\Delta M_{B_s}$.

Before closing this section, let us comment on the charm-quark contribution. 
In the analysis, we focused on the top-quark contributions in the box diagrams and kept the charm-quark ones aside. 
This approximation is appropriate in the $B_{s,d}$ meson system.
However, they are dominant in the $K$ meson system, e.g., for $\epsilon_K$ in the left-right symmetric model~\cite{Bertolini:2014sua}.
Then, the SMEFT and charged Higgs corrections explored in this paper become necessary, and long-distance effects should be taken into account.
This topic will be studied in  the future.

\section{Conclusions}

Since the experimental constraints push the NP scale higher, the NP particle masses are likely to be much larger than the SM ones, i.e., the EWSB scale. 
Then, FCNC amplitudes should be investigated in the framework of the SMEFT rather than the ``low-scale'' one. 
In a class of the NP models, both of the NP and SM particles contribute to a loop diagram simultaneously. 
In order to reduce the uncertainty of the input scale of the Wilson coefficients particularly in such models, we studied the SMEFT corrections, paying attention to the top-quark effects.
For the FCNC observables, the operator matching needs to be performed at the one-loop level. 
We provide the complete one-loop matching formula for $\Delta F=2$ transitions at $\mathcal{O}(y_t^2)$. 

We also investigated $\Delta M_{B_s}$ in the left-right symmetric models.
The right-handed $W$ boson generates the flavor transitions similarly to the left-handed one in the SM.
The SMEFT corrections are studied and compared with the conventional results. 
We found that the Wilson coefficient $C_{4}$ is affected by $\mathcal{O}(1)\%$ and $C_{5}$ by $\mathcal{O}(100)\%$.
Since the observable $\Delta M_{B_s}$ is dominated by the former quantitatively, the SMEFT effects for $\Delta M_{B_s}$ become comparable to the result in $C_{4}$.
In addition to the SMEFT effects, we discussed the contribution of the heavy charged Higgs boson.
Although it can be comparable to the one-loop corrections to the heavy neutral Higgs boson contribution, which are necessary for the gauge invariance, the effect has often been neglected in the literature. 
It was found that the relative contribution is about one percent level and almost independent of $M_{W_{R}}$.

Although the difference between our and conventional results becomes smaller if $\mu = m_t$ is chosen for $\Delta M_{B_s}$ in the left-right symmetric models, the deviation becomes enhanced as $\mu_{LR}$ increases. 
In order to clarify in which energy scale the Wilson coefficients should be input, it is important to take account of the SMEFT RGEs and matching conditions for the NP models in high scales.

\vspace{1em}
\noindent {\it Acknowledgements}:
We are grateful to Toru Goto, Satoshi Mishima and Kei Yamamoto for the collaboration in the early stage of this project. 
We also thank Jason Aebischer, Andreas Crivellin and Matteo Fael for discussions. 
D.U. would like to thank Matthias Steinhauser and all the members of TTP at Karlsruhe Institute of Technology for kind hospitality during the stay. 
This work was supported by JSPS KAKENHI No.~16K17681 (M.E.) and 16H03991 (M.E.).

\appendix
\section{Renormalization group equations}
\label{sec:RGE}

In this appendix, we summarize the SMEFT RGEs which are relevant for the $\Delta F=2$ observables at the one-loop level.
We focus on the anomalous dimensions which depend on the top-Yukawa or QCD couplings. 
In the following expressions, we define 
\begin{align}
 \dot{C}_a \equiv (4\pi)^2 \frac{d C_a}{d\ln\mu},~~~
 X_t \equiv \frac{\pi\alpha}{s_W^2} x_t.
\end{align}
The anomalous dimensions at $\mathcal{O}(y_t^2)$ and $\mathcal{O}(g_s^2)$ are obtained as (see Refs.~\cite{Jenkins:2013zja,Jenkins:2013wua,Alonso:2013hga} for the complete one-loop formula of the SMEFT RGEs):
\begin{align}
 (\dot{C}^{(1)}_{Hq})_{pr} &= 
 X_t \bigg[
 \lambda_t^{pr} (C_{H\Box} + C_{HD}) 
 - 2 \lambda_t^{pr} (C_{Hu})_{33} 
 + 3 \lambda_t^{pt} (C^{(1)}_{Hq})_{tr} 
 + 3 \lambda_t^{tr} (C^{(1)}_{Hq})_{pt}
 \notag \\ & \quad\quad
 - 9 \lambda_t^{pt} (C^{(3)}_{Hq})_{tr} 
 - 9 \lambda_t^{tr} (C^{(3)}_{Hq})_{pt}
 \notag \\ & \quad\quad
 + 2 \lambda_t^{ts} \Big(
     6 (C^{(1)}_{qq})_{prst} 
   + 6 (C^{(1)}_{qq})_{stpr} 
   + (C^{(1)}_{qq})_{ptsr} 
   + (C^{(1)}_{qq})_{srpt}
   + 3 (C^{(3)}_{qq})_{ptsr} 
   + 3 (C^{(3)}_{qq})_{srpt} \Big)
 \notag \\ & \quad\quad
 - 12 \lambda_t^{kk} (C^{(1)}_{qu})_{pr33} 
 + 12 \lambda_t^{kk} (C^{(1)}_{Hq})_{pr} 
 + \lambda_t^{pt} (C^{(1)}_{Hq})_{tr} 
 + \lambda_t^{tr} (C^{(1)}_{Hq})_{pt}
 \bigg],
 \\
 (\dot{C}^{(3)}_{Hq})_{pr} &= 
 X_t \bigg[
 - \lambda_t^{pr} C_{H\Box} 
 - 3 \lambda_t^{pt} (C^{(1)}_{Hq})_{tr} 
 - 3 \lambda_t^{tr} (C^{(1)}_{Hq})_{pt}
 + \lambda_t^{pt} (C^{(3)}_{Hq})_{tr} 
 + \lambda_t^{tr} (C^{(3)}_{Hq})_{pt}
 \notag \\ & \quad\quad
 - 2 \lambda_t^{ts} \Big(
    6 (C^{(3)}_{qq})_{prst} 
  + 6 (C^{(3)}_{qq})_{stpr} 
  + (C^{(1)}_{qq})_{ptsr} 
  + (C^{(1)}_{qq})_{srpt} 
  - (C^{(3)}_{qq})_{ptsr} 
  - (C^{(3)}_{qq})_{srpt} \Big)
 \notag \\ & \quad\quad
 + 12 \lambda_t^{kk} (C^{(3)}_{Hq})_{pr}
 + \lambda_t^{pt} (C^{(3)}_{Hq})_{tr} 
 + \lambda_t^{tr} (C^{(3)}_{Hq})_{pt} \bigg],
 \\
 (\dot{C}_{Hd})_{pr} &=
 X_t \bigg[ 
 - 12 \lambda_t^{kk} (C^{(1)}_{ud})_{33pr} 
 + 12 \lambda_t^{ts} (C^{(1)}_{qd})_{stpr} 
 + 12 \lambda_t^{kk} (C_{Hd})_{pr} \bigg],
 \\
 (\dot{C}_{Hu})_{pr} &=
 X_t \bigg[ 
 - 2 \lambda_t^{kk} \delta_{p3} \delta_{3r} (C_{H\Box}+C_{HD})
 - 4 \lambda_t^{ts} \delta_{p3} \delta_{3r} (C^{(1)}_{Hq})_{st}
 + 6 \lambda_t^{kk} \delta_{p3} (C_{Hu})_{3r}
 + 6 \lambda_t^{kk} \delta_{3r} (C_{Hu})_{p3}
 \notag \\ & \quad\quad
 - 4 \lambda_t^{kk} \Big( 
    3 (C_{uu})_{pr33}
  + 3 (C_{uu})_{33pr}
  +   (C_{uu})_{p33r}
  +   (C_{uu})_{3rp3} \Big)
 \notag \\ & \quad\quad
 + 12 \lambda_t^{ts} (C^{(1)}_{qu})_{stpr}
 + 12 \lambda_t^{kk} (C_{Hu})_{pr}
 + 2 \lambda_t^{kk} \delta_{p3} (C_{Hu})_{3r}
 + 2 \lambda_t^{kk} \delta_{r3} (C_{Hu})_{p3} \bigg],
 \\
 \dot{C}_{H\Box} &=
 X_t \bigg[
 - 2 \Big(
  -  6 \lambda_t^{sr} (C^{(1)}_{Hq})_{rs}
  + 18 \lambda_t^{sr} (C^{3}_{Hq})_{rs}
  +  6 \lambda_t^{kk} (C_{Hu})_{33} \Big)
 + 24 \lambda_t^{kk} C_{H\Box} \bigg],
 \\
 \dot{C}_{HD} &=
 X_t \bigg[
 - 2 \Big(
  - 24 \lambda_t^{sr} (C^{(1)}_{Hq})_{rs}
  + 24 \lambda_t^{kk} (C_{Hu})_{33} \Big)
 + 24 \lambda_t^{kk} C_{HD} \bigg],
 \\
 (\dot{C}_{uu})_{prst} &=
 X_t \bigg[
 - 2 \lambda_t^{kk} \delta_{p3} \delta_{3r} (C_{Hu})_{st}
 - 2 \lambda_t^{kk} \delta_{s3} \delta_{3t} (C_{Hu})_{pr} 
 \notag \\ & \quad\quad
 - 2 \lambda_t^{wv} \delta_{r3} \delta_{p3} (C^{(1)}_{qu})_{vwst}
 - 2 \lambda_t^{wv} \delta_{t3} \delta_{s3} (C^{(1)}_{qu})_{vwpr}
 \notag \\ & \quad\quad
 + \frac{1}{3} \lambda_t^{wv} \delta_{p3} \delta_{r3} (C^{(8)}_{qu})_{vwst}
 + \frac{1}{3} \lambda_t^{wv} \delta_{s3} \delta_{t3} (C^{(8)}_{qu})_{vwpr}
 \notag \\ & \quad\quad
 - \lambda_t^{wv} \delta_{s3} \delta_{3r} (C^{(8)}_{qu})_{vwpt}
 - \lambda_t^{wv} \delta_{t3} \delta_{p3} (C^{(8)}_{qu})_{vwsr}
 \notag \\ & \quad\quad
 + 2 \lambda_t^{kk} \delta_{p3} (C_{uu})_{3rst} 
 + 2 \lambda_t^{kk} \delta_{s3} (C_{uu})_{pr3t}
 + 2 \lambda_t^{kk} \delta_{r3} (C_{uu})_{p3st}
 + 2 \lambda_t^{kk} \delta_{t3} (C_{uu})_{prs3} \bigg]
 \notag \\ & 
 + 4\pi \alpha_s \bigg[ 
   \frac{1}{3} (C^{(8)}_{qu})_{wwpt} \delta_{rs} 
 + \frac{1}{3} (C^{(8)}_{qu})_{wwsr} \delta_{pt} 
 - \frac{1}{3N_c} (C^{(8)}_{qu})_{wwst} \delta_{pr} 
 - \frac{1}{3N_c} (C^{(8)}_{qu})_{wwpr} \delta_{st} 
 \notag \\ & \quad\quad\quad
 + \frac{1}{3} (C_{uu})_{pwwt} \delta_{rs}
 + \frac{1}{3} (C_{uu})_{swwr} \delta_{pt} 
 + \frac{1}{3} (C_{uu})_{wtpw} \delta_{rs}
 + \frac{1}{3} (C_{uu})_{wrsw} \delta_{pt}
 \notag \\ & \quad\quad\quad
 - \frac{1}{3N_c} (C_{uu})_{pwwr} \delta_{st}
 - \frac{1}{3N_c} (C_{uu})_{swwt} \delta_{pr}
 - \frac{1}{3N_c} (C_{uu})_{wrpw} \delta_{st}
 \notag \\ & \quad\quad\quad
 - \frac{1}{3N_c} (C_{uu})_{wtsw} \delta_{pr}
 + \frac{1}{6} (C^{(8)}_{ud})_{ptww} \delta_{rs}
 + \frac{1}{6} (C^{(8)}_{ud})_{srww} \delta_{pt} 
 - \frac{1}{6N_c} (C^{(8)}_{ud})_{prww} \delta_{st}
 \notag \\ & \quad\quad\quad
 - \frac{1}{6N_c} (C^{ (8)}_{ud})_{stww} \delta_{pr}
 + 6 (C_{uu})_{ptsr}
 - \frac{6}{N_c}(C_{uu})_{prst} \bigg],
 \\
 (\dot{C}^{(1)}_{ud})_{prst} &=
 X_t \bigg[
 - 4 \lambda_t^{kk} \delta_{p3} \delta_{3r} (C_{Hd})_{st} 
 - 4 \lambda_t^{wv} \delta_{p3} \delta_{r3} (C^{(1)}_{qd})_{vwst}
 + 2 \lambda_t^{kk} \delta_{p3} (C^{(1)}_{ud})_{3rst}
 + 2 \lambda_t^{kk} \delta_{r3} (C^{(1)}_{ud})_{p3st} \bigg]
 \notag \\ & 
 + 3 \left( \frac{N_c^2-1}{N_c^2} \right) 4\pi \alpha_s (C^{(8)}_{ud})_{prst},
 \\
 (\dot{C}^{(8)}_{ud})_{prst} &=
 X_t \bigg[
 - 4 \lambda_t^{wv} \delta_{r3} \delta_{p3} (C^{(8)}_{qd})_{vwst}
 + 2 \lambda_t^{kk} \delta_{p3} (C^{(8)}_{ud})_{3rst}
 + 2 \lambda_t^{kk} \delta_{r3} (C^{(8)}_{ud})_{p3st} \bigg]
 \notag \\ & 
 + 4\pi \alpha_s \bigg[
   \frac{4}{3} (C_{uu})_{pwwr} \delta_{st} 
 + \frac{4}{3} (C_{uu})_{wrpw} \delta_{st} 
 + \frac{4}{3} (C_{dd})_{swwt} \delta_{pr}
 + \frac{4}{3} (C_{dd})_{wtsw} \delta_{pr} 
 \notag \\ & \quad\quad\quad
 + \frac{4}{3} (C^{(8)}_{qu})_{wwpr} \delta_{st}
 + \frac{4}{3} (C^{(8)}_{qd})_{wwst} \delta_{pr} 
 \notag \\ & \quad\quad\quad
 + \frac{2}{3} (C^{(8)}_{ud})_{prww} \delta_{st} 
 + \frac{2}{3} (C^{(8)}_{ud})_{wwst} \delta_{pr} 
 - 12 \frac{1}{N_c} (C^{(8)}_{ud})_{prst} 
 + 12               (C^{(1)}_{ud})_{prst} \bigg],
  \\
 (\dot{C}^{(1)}_{qu})_{prst} &=
 X_t \bigg[ 
   \frac{4}{3} \lambda_t^{wr} \delta_{s3} \Big(
                (C^{(1)}_{qu})_{pw3t} 
  + \frac{4}{3} (C^{(8)}_{qu})_{pw3t} \Big)
 + \frac{4}{3} \lambda_t^{pv} \delta_{t3} \Big(
                (C^{(1)}_{qu})_{rv3s}^*
  + \frac{4}{3} (C^{(8)}_{qu})_{rv3s}^* \Big)
 \notag \\ & \quad\quad
 + 2 \lambda_t^{pr} (C_{Hu})_{st} 
 - 4 \lambda_t^{kk} \delta_{s3} \delta_{3t} (C^{(1)}_{Hq})_{pr}
 + \frac{1}{3} \Big(
    2 \lambda_t^{pv} \delta_{s3} (C^{(1)}_{qu})_{vr3t}
  + 2 \lambda_t^{vr} \delta_{3t} (C^{(1)}_{qu})_{pvs3} \Big)
 \notag \\ & \quad\quad
 - \frac{1}{9} \Big(
    \lambda_t^{pv} \delta_{s3} (C^{(8)}_{qu})_{vr3t}
  + \lambda_t^{vr} \delta_{3t} (C^{(8)}_{qu})_{pvs3} \Big)
 \notag \\ & \quad\quad
 - \frac{2}{3} \Big(
    \lambda_t^{vw} \delta_{s3} \delta_{3t} (C^{(1)}_{qq})_{pvwr}
  + \lambda_t^{wv} \delta_{s3} \delta_{3t} (C^{(1)}_{qq})_{pwvr}
  + \lambda_t^{pr} (C_{uu})_{3ts3}
  + \lambda_t^{pr} (C_{uu})_{3ts3} \Big)
 \notag \\ & \quad\quad
 - 2 \Big(
    \lambda_t^{vw} \delta_{s3} \delta_{3t} (C^{(3)}_{qq})_{pvwr}
  + \lambda_t^{wv} \delta_{s3} \delta_{3t} (C^{(3)}_{qq})_{pwvr} \Big)
 \notag \\ & \quad\quad
 + \Big(
    \lambda_t^{pv} \delta_{s3} (C^{(8)}_{qu})_{vr3t}
  + \lambda_t^{vr} \delta_{3t} (C^{(8)}_{qu})_{pvs3} \Big)
 - 8 \lambda_t^{wv} \delta_{s3} \delta_{3t} (C^{(1)}_{qq})_{prvw}
 - 4 \lambda_t^{pr} (C_{uu})_{33st}
 \notag \\ & \quad\quad
 +   \lambda_t^{pv} (C^{(1)}_{qu})_{vrst}
 + 2 \lambda_t^{kk} \delta_{s3} (C^{(1)}_{qu})_{pr3t} 
 +   \lambda_t^{vr} (C^{(1)}_{qu})_{pvst} 
 + 2 \lambda_t^{kk} \delta_{t3} (C^{(1)}_{qu})_{prs3} \bigg]
 \notag \\ & 
 - 3 \left( \frac{N_c^2 -1}{N_c^2} \right) 4\pi \alpha_s (C^{(8)}_{qu})_{prst},
 \\
 (\dot{C}^{(8)}_{qu})_{prst} &=
 X_t \bigg[ 
   8 \lambda_t^{wr} \delta_{s3} \Big(
                (C^{(1)}_{qu})_{pw3t}
  + \frac{4}{3} (C^{(8)}_{qu})_{pw3t} \Big)
 + 8 \lambda_t^{pv} \delta_{t3} \Big(
                (C^{(1)}_{qu})_{rv3s}^*
  + \frac{4}{3} (C^{(8)}_{qu})_{rv3s}^* \Big)
 \notag \\ & \quad\quad
 - \frac{2}{3} \Big(
    \lambda_t^{pv} \delta_{s3} (C^{(8)}_{qu})_{vr3t}
  + \lambda_t^{vr} \delta_{3t} (C^{(8)}_{qu})_{pvs3} \Big)
 \notag \\ & \quad\quad
 - 4 \Big(
    \lambda_t^{vw} \delta_{s3} \delta_{3t} (C^{(1)}_{qq})_{pvwr} 
  + \lambda_t^{wv} \delta_{s3} \delta_{3t} (C^{(1)}_{qq})_{pwvr}
  - \lambda_t^{pv} \delta_{s3} (C^{(1)}_{qu})_{vr3t}
  - \lambda_t^{vr} \delta_{3t} (C^{(1)}_{qu})_{pvs3} \Big)
 \notag \\ & \quad\quad
 - 4 \Big(
    \lambda_t^{pr} (C_{uu})_{3ts3}
  + \lambda_t^{pr} (C_{uu})_{3ts3} \Big)
 - 12 \Big(
    \lambda_t^{vw} \delta_{s3} \delta_{3t} (C^{(3)}_{qq})_{pvwr}
  + \lambda_t^{wr} \delta_{s3} \delta_{3t} (C^{(3)}_{qq})_{pwvr} \Big)
 \notag \\ & \quad\quad
 +   \lambda_t^{pv} (C^{(8)}_{qu})_{vrst}
 + 2 \lambda_t^{kk} \delta_{s3} (C^{(8)}_{qu})_{pr3t}
 +   \lambda_t^{vr} (C^{(8)}_{qu})_{pvst}
 + 2 \lambda_t^{kk} \delta_{t3} (C^{(8)}_{qu})_{prs3} \bigg]
 \notag \\ & 
 + 4\pi \alpha_s \bigg[
   \frac{4}{3} (C^{(1)}_{qq})_{pwwr} \delta_{st}
 + \frac{4}{3} (C^{(1)}_{qq})_{wrpw} \delta_{st} 
 + 4 (C^{(3)}_{qq})_{pwwr} \delta_{st}
 + 4 (C^{(3)}_{qq})_{wrpw} \delta_{st} 
 \notag \\ & \quad\quad\quad
 + \frac{2}{3} (C^{(8)}_{qu})_{prww} \delta_{st}
 + \frac{2}{3} (C^{(8)}_{qd})_{prww} \delta_{st} 
 + \frac{4}{3} (C^{(8)}_{qu})_{wwst} \delta_{pr}
 \notag \\ & \quad\quad\quad
 + \frac{2}{3} (C^{(8)}_{ud})_{stww} \delta_{pr}
 + \frac{4}{3} (C_{uu})_{swwt} \delta_{pr}
 + \frac{4}{3} (C_{uu})_{wtsw} \delta_{pr} 
 \notag \\ & \quad\quad\quad
 - 6 \left(N_c-\frac{2}{N_c}\right) (C^{(8)}_{qu})_{prst}
 - 12 (C^{(1)}_{qu})_{prst} \bigg],
 \\
 (\dot{C}^{(1)}_{qq})_{prst} &=
 X_t \bigg[
   \lambda_t^{pr} (C^{(1)}_{Hq})_{st}
 + \lambda_t^{st} (C^{(1)}_{Hq})_{pr}
 \notag \\ & \quad\quad
 + \frac{1}{6} \Big(
    \lambda_t^{pr} (C^{(8)}_{qu})_{st33}
  + \lambda_t^{st} (C^{(8)}_{qu})_{pr33} \Big)
 - \frac{1}{4} \Big(
    \lambda_t^{pt} (C^{(8)}_{qu})_{sr33}
  + \lambda_t^{sr} (C^{(8)}_{qu})_{pt33} \Big)
 \notag \\ & \quad\quad
 - \lambda_t^{pr} (C^{(1)}_{qu})_{st33}
 - \lambda_t^{st} (C^{(1)}_{qu})_{pr33}
 \notag \\ & \quad\quad
 + \lambda_t^{pv} (C^{(1)}_{qq})_{vrst}
 + \lambda_t^{sv} (C^{(1)}_{qq})_{prvt}
 + \lambda_t^{vr} (C^{(1)}_{qq})_{pvst} 
 + \lambda_t^{vt} (C^{(1)}_{qq})_{prsv} \bigg]
 \notag \\ & 
 + 4\pi \alpha_s \bigg[
   3 (C^{(1)}_{qq})_{ptsr}
 + 9 (C^{(3)}_{qq})_{ptsr}
 - \frac{6}{N_c} (C^{(1)}_{qq})_{prst}
 \notag \\ & \quad\quad\quad
 + \frac{1}{6} (C^{(1)}_{qq})_{swwr} \delta_{pt}
 + \frac{1}{6} (C^{(1)}_{qq})_{pwwt} \delta_{rs}
 + \frac{1}{6} (C^{(1)}_{qq})_{wrsw} \delta_{pt}
 + \frac{1}{6} (C^{(1)}_{qq})_{wtpw} \delta_{rs}
 \notag \\ & \quad\quad\quad
 - \frac{1}{3N_c} (C^{(1)}_{qq})_{pwwr} \delta_{st}
 - \frac{1}{3N_c} (C^{(1)}_{qq})_{swwt} \delta_{pr}
 - \frac{1}{3N_c} (C^{(1)}_{qq})_{wrpw} \delta_{st} 
 - \frac{1}{3N_c} (C^{(1)}_{qq})_{wtsw} \delta_{pr}
 \notag \\ & \quad\quad\quad
 + \frac{1}{2} (C^{(3)}_{qq})_{swwr} \delta_{pt}
 + \frac{1}{2} (C^{(3)}_{qq})_{pwwt} \delta_{rs}
 + \frac{1}{2} (C^{(3)}_{qq})_{wrsw} \delta_{pt}
 + \frac{1}{2} (C^{(3)}_{qq})_{wtpw} \delta_{rs} 
 \notag \\ & \quad\quad\quad
 - \frac{1}{N_c} (C^{(3)}_{qq})_{pwwr} \delta_{st}
 - \frac{1}{N_c} (C^{(3)}_{qq})_{swwt} \delta_{pr}
 - \frac{1}{N_c} (C^{(3)}_{qq})_{wrpw} \delta_{st}
 - \frac{1}{N_c} (C^{(3)}_{qq})_{wtsw} \delta_{pr}
 \notag \\ & \quad\quad\quad
 + \frac{1}{12}   (C^{(8)}_{qu})_{srww} \delta_{pt} 
 + \frac{1}{12}   (C^{(8)}_{qu})_{ptww} \delta_{rs}
 - \frac{1}{6N_c} (C^{(8)}_{qu})_{prww} \delta_{st}
 - \frac{1}{6N_c} (C^{(8)}_{qu})_{stww} \delta_{pr} 
 \notag \\ & \quad\quad\quad
 + \frac{1}{12}   (C^{(8)}_{qd})_{srww} \delta_{pt}
 + \frac{1}{12}   (C^{(8)}_{qd})_{ptww} \delta_{rs}
 - \frac{1}{6N_c} (C^{(8)}_{qd})_{prww} \delta_{st}
 - \frac{1}{6N_c} (C^{(8)}_{qd})_{stww} \delta_{pr} \bigg],
 \\
 (\dot{C}^{(3)}_{qq})_{prst} &=
 X_t \bigg[
 - \lambda_t^{pr} (C^{(3)}_{Hq})_{st}
 - \lambda_t^{st} (C^{(3)}_{Hq})_{pr}
 - \frac{1}{4} \Big(
    \lambda_t^{pt} (C^{(8)}_{qu})_{sr33}
  + \lambda_t^{sr} (C^{(8)}_{qu})_{pt33} \Big)
 \notag \\ & \quad\quad
 + \lambda_t^{pv} (C^{(3)}_{qq})_{vrst}
 + \lambda_t^{sv} (C^{(3)}_{qq})_{prvt}
 + \lambda_t^{vr} (C^{(3)}_{qq})_{pvst}
 + \lambda_t^{vt} (C^{(3)}_{qq})_{prsv} \bigg]
 \notag \\ & 
 + 4\pi \alpha_s \bigg[
 - 3             (C^{(3)}_{qq})_{ptsr}
 - \frac{6}{N_c} (C^{(3)}_{qq})_{prst}
 + 3             (C^{(1)}_{qq})_{ptsr}
 \notag \\ & \quad\quad\quad
 + \frac{1}{6} (C^{(1)}_{qq})_{pwwt} \delta_{rs}
 + \frac{1}{6} (C^{(1)}_{qq})_{swwr} \delta_{pt} 
 + \frac{1}{6} (C^{(1)}_{qq})_{wtpw} \delta_{rs} 
 + \frac{1}{6} (C^{(1)}_{qq})_{wrsw} \delta_{pt}
 \notag \\ & \quad\quad\quad
 + \frac{1}{2} (C^{(3)}_{qq})_{pwwt} \delta_{sr}
 + \frac{1}{2} (C^{(3)}_{qq})_{swwr} \delta_{pt}
 + \frac{1}{2} (C^{(3)}_{qq})_{wtpw} \delta_{rs}
 + \frac{1}{2} (C^{(3)}_{qq})_{wrsw} \delta_{pt}
 \notag \\ & \quad\quad\quad
 + \frac{1}{12} (C^{(8)}_{qu})_{ptww} \delta_{rs}
 + \frac{1}{12} (C^{(8)}_{qu})_{srww} \delta_{pt} 
 + \frac{1}{12} (C^{(8)}_{qd})_{ptww} \delta_{rs}
 + \frac{1}{12} (C^{(8)}_{qd})_{srww} \delta_{pt} \bigg],
 \\
 (\dot{C}_{dd})_{prst} &=
 4\pi \alpha_s \bigg[
   6             (C_{dd})_{ptsr}
 - \frac{6}{N_c} (C_{dd})_{prst}
 \notag \\ & \quad\quad\quad
 + \frac{1}{3} (C_{dd})_{pwwt} \delta_{rs}
 + \frac{1}{3} (C_{dd})_{wtpw} \delta_{rs} 
 + \frac{1}{3} (C_{dd})_{wrsw} \delta_{pt} 
 - \frac{1}{3N_c} (C_{dd})_{pwwr} \delta_{st}
 \notag \\ & \quad\quad\quad
 - \frac{1}{3N_c} (C_{dd})_{swwt} \delta_{pr}
 - \frac{1}{3N_c} (C_{dd})_{wtsw} \delta_{pr}
 - \frac{1}{3N_c} (C_{dd})_{wrpw} \delta_{st}
 \notag \\ & \quad\quad\quad
 + \frac{1}{3}    (C^{(8)}_{qd})_{wwsr} \delta_{pt}
 + \frac{1}{3}    (C^{(8)}_{qd})_{wwpt} \delta_{rs} 
 - \frac{1}{3N_c} (C^{(8)}_{qd})_{wwpr} \delta_{st}
 - \frac{1}{3N_c} (C^{(8)}_{qd})_{wwst} \delta_{pr}
 \notag \\ & \quad\quad\quad
 + \frac{1}{6}    (C^{(8)}_{ud})_{wwpt} \delta_{rs} 
 + \frac{1}{6}    (C^{(8)}_{ud})_{wwsr} \delta_{pt}
 - \frac{1}{6N_c} (C^{(8)}_{ud})_{wwpr} \delta_{st}
 - \frac{1}{6N_c} (C^{(8)}_{ud})_{wwst} \delta_{pr} \bigg],
 \\
 (\dot{C}^{(1)}_{qd})_{prst} &=
 X_t \bigg[
   2 \lambda_t^{pr} (C_{Hd})_{st} 
 - 2 \lambda_t^{pr} (C^{(1)}_{ud})_{33st}
 +   \lambda_t^{pv} (C^{(1)}_{qd})_{vrst}
 +   \lambda_t^{vr} (C^{(1)}_{qd})_{pvst} \bigg]
 \notag \\ & 
 - 3 \left(\frac{N_c^2-1}{N_c^2}\right) 4\pi \alpha_s (C^{(8)}_{qd})_{prst},
 \\
 (\dot{ C}^{(8)}_{qd})_{prst} &=
 X_t \bigg[
 - 2 \lambda_t^{pr} (C^{(8)}_{ud})_{33st}
 +   \lambda_t^{pv} (C^{(8)}_{qd})_{vrst}
 +   \lambda_t^{vr} ( C^{(8)}_{qd})_{pvst} \bigg]
 \notag \\ & 
 + 4\pi \alpha_s \bigg[
   \frac{4}{3} (C^{(1)}_{qq})_{pwwr} \delta_{st} 
 + \frac{4}{3} (C^{(1)}_{qq})_{wrpw} \delta_{st} 
 + 4 (C^{(3)}_{qq})_{pwwr} \delta_{st} 
 + 4 (C^{(3)}_{qq})_{wrpw} \delta_{st}
 \notag \\ & \quad\quad\quad
 + \frac{2}{3} (C^{(8)}_{qu})_{prww} \delta_{st}
 + \frac{2}{3} (C^{(8)}_{qd})_{prww} \delta_{st}
 + \frac{4}{3} (C^{(8)}_{qd})_{wwst} \delta_{pr}
 + \frac{2}{3} (C^{(8)}_{ud})_{wwst} \delta_{pr} 
 \notag \\ & \quad\quad\quad
 + \frac{4}{3} (C_{dd})_{swwt} \delta_{pr}
 + \frac{4}{3} (C_{dd})_{wtsw} \delta_{pr}
 - 6 \left(N_c -\frac{2}{N_c}\right) (C^{(8)}_{qd})_{prst}
 - 12 (C^{(1)}_{qd})_{prst} \bigg].
\end{align}

\section{Higgs sector in left-right symmetric models}
\label{sec:Higgssector}

In this section, we briefly review the Higgs sector in the left-right symmetric models.
After the left-right symmetry is broken, the scalar potential with $v_L =0$ \cite{Deshpande:1990ip} is 
\begin{align}
 V =\,& 
 - \mu_1^2 \textrm{Tr}\left( \Phi^{\dag} \Phi \right)
 - \mu_2^2 \left[ \textrm{Tr}\left( \tilde{\Phi} \Phi^{\dag} \right) 
 + \textrm{Tr}\left( \tilde{\Phi}^{\dag} \Phi  \right) \right] 
 \notag \\ & 
 + \lambda_1 \left[ \textrm{Tr}\left( \Phi^{\dag} \Phi \right)\right]^2
 + \lambda_2 \left\{ \left[ \textrm{Tr} \left( \tilde{\Phi} \Phi^{\dag}\right) \right]^2 
 + \left[ \textrm{Tr} \left( \tilde{\Phi}^{\dag} \Phi \right) \right]^2 \right\} 
 \notag \\ & 
 + \lambda_3 \textrm{Tr} \left( \tilde{\Phi} \Phi^{\dag} \right) 
             \textrm{Tr} \left( \tilde{\Phi}^{\dag} \Phi \right) 
 + \lambda_4 \textrm{Tr} \left( \Phi^{\dag} \Phi \right) 
 \left[ 
   \textrm{Tr} \left( \tilde{\Phi} \Phi^{\dag} \right) 
 + \textrm{Tr} \left( \tilde{\Phi}^{\dag} \Phi \right) \right] 
 \notag \\ & 
 + \alpha_1 \textrm{Tr} \left( \Phi^{\dag} \Phi \right) 
            \textrm{Tr} \left(\langle \Delta _R^{\dag} \rangle \langle\Delta_R \rangle \right)
 + \alpha_2 \left[ e^{  i \delta} \textrm{Tr} \left( \tilde{\Phi}^{\dag} \Phi \right)
                 + e^{- i \delta} \textrm{Tr} \left( \tilde{\Phi} \Phi^{\dag} \right) \right] 
 \notag \\ & 
 + \alpha_3 \textrm{Tr} 
 \left( \Phi^{\dag} \Phi \langle \Delta_R \rangle \langle\Delta_R^{\dag} \rangle \right),
\end{align}
where $\tilde{\Phi} = \sigma_2 \Phi^{\ast} \sigma_2$.
Under this scalar potential, the Higgs bi-doublet $\Phi$ obtains complex VEVs as Eq.~\eqref{eq:higgsVEV} and  the spontaneous $CP$-violating phase $\alpha$ emerges at the EWSB vacuum.

In the limit of $v_R \gg v$, the following linear combinations diagonalize the neutral and charged Higgs mass matrices,
\begin{align}
 H^0 &= \cos \beta \phi_2^0 - \sin \beta  e^{ i \alpha}\phi_1^{0 \ast}, \\
 h^0 &= \sin \beta e^{- i \alpha} \phi_2^0 + \cos \beta \phi_1^{0\ast}, \\
 H^+ &= \cos \beta \phi_2^+ +  \sin \beta e^{ i \alpha} \phi_1^{+}, \\
 G^+ &= \sin \beta \phi_2^+ - \cos \beta   e^{ i \alpha}  \phi_1^{+},
\end{align}
where $H^0$ ($H^+$) is the heavy neutral (charged) Higgs, $G^+$ the NG boson, and $h^0$ includes SM Higgs and NG boson components.
The heavy Higgs masses are obtained as
\begin{align}
 M_{H^{0}}^2 = M_{H^{\pm}}^2 = \frac{\alpha_3 v_R^2}{2 \cos 2 \beta} \equiv M_H^2.
\end{align}

The Yukawa interactions in the gauge eigenstate basis are
\begin{align}
 - \mathcal{L}_Y &=
 \overline{Q}_L \left(Y \Phi + \tilde{Y} \tilde{\Phi} \right) Q_R + \textrm{h.c.}
 \notag \\ &\supset 
   \overline{U}_L S_L^u M_U S_R^{u \dagger} U_R 
 + \overline{D}_L S_L^d M_D S_R^{d \dagger} D_R + \textrm{h.c.}
 \notag \\ &\equiv 
   \overline{u}_L M_U u_R 
 + \overline{d}_L M_D d_R + \textrm{h.c.},
\end{align}
with the mass matrices,
\begin{align}
 S_L^u M_U S_R^{u \dagger} &= \frac{v}{\sqrt{2}}
 \left( Y \cos \beta               + \tilde{Y} \sin \beta e^{- i \alpha} \right) ,
 \label{eq:YYt1} \\
 S_L^d M_D S_R^{d \dagger} &= \frac{v}{\sqrt{2}}
 \left( Y \sin \beta e^{ i \alpha} + \tilde{Y} \cos \beta  \right).
 \label{eq:YYt2}
\end{align}
Here, $u_{L,R}$ and $d_{L,R}$ represent the quark mass eigenstates with $M_U = \textrm{diag}(m_u, m_c, m_t)$ and $M_D = \textrm{diag}(m_d, m_s, m_b)$. 
The unitary matrices $S_{L,R}^{u,d}$ satisfy
\begin{align}
V_L = S_ L^{u \dag} S_ L^d,~~~
V_R = S_ R^{u \dag} S_ R^d.
\end{align} 
From Eq.~\eqref{eq:YYt1} and \eqref{eq:YYt2}, $Y$ and $\tilde{Y}$ are written as
 \begin{align}
 Y &= 
 \frac{\sqrt{2}}{v \cos 2 \beta}\left( 
   \cos \beta S_L^u M_U S_R^{u \dagger} 
 - \sin \beta e^{- i \alpha} S_L^d M_D S_R^{d \dagger} \right),\notag \\
 \tilde{Y} &= 
 \frac{\sqrt{2}}{v \cos 2 \beta}\left( 
 - \sin \beta e^{  i \alpha} S_L^u M_U S_R^{u \dagger} 
 + \cos \beta S_L^d M_D S_R^{d \dagger} \right).
 \label{eq:YYtsol}
\end{align}
Then, the Yukawa interactions are represented in the mass eigenstate basis as
\begin{align}
 -\mathcal{L}_Y =\,&
 \overline{Q}_L \left(Y \Phi + \tilde{Y} \tilde{\Phi} \right) Q_R + \textrm{h.c.}
 \notag \\ =\,& 
 \overline{u}_L S_L^{u \dagger} 
 \left( Y \phi_1^0 + \tilde{Y} \phi_2^{0 \ast} \right)
 S_R^u u_R 
 + 
 \overline{d}_L S_L^{d \dagger} 
 \left( Y \phi_2^0 + \tilde{Y} \phi_1^{0 \ast} \right)
 S_R^d d_R 
 \notag \\ & 
 + \overline{u}_L S_L^{u \dagger} \left( Y \phi_2^+ - \tilde{Y} \phi_1^{+} \right) S_R^{d} d_R
 + \overline{d}_L S_L^{d \dagger} \left( Y \phi_1^- - \tilde{Y} \phi_2^{-} \right) S_R^{u} u_R
 + \textrm{h.c.}
 \notag \\ =\,& 
   \frac{\sqrt{2}}{v} \overline{u}_L S_L^{u \dagger} \left[ 
 S_L^u M_U S_R^{u \dagger} h^{\ast} 
 + \frac{1}{\cos 2 \beta} \left( 
 S_L^d M_D S_R^{d \dagger} 
 - \sin 2 \beta e^{ i \alpha} 
 S_L^u M_U S_R^{u \dagger} \right) H^{0\ast} \right] S_R^u u_R 
 \notag \\ &
 + \frac{\sqrt{2}}{v} \overline{d}_L S_L^{d \dagger} \left[ 
 S_ L^d M_D S_R^{d \dagger} h 
 + \frac{1}{\cos 2 \beta} \left(
 S_L^u M_U S_R^{u \dagger} 
 - \sin 2 \beta e^{-i \alpha} 
 S_ L^d M_D S_R^{d \dagger} \right) H^0 \right] S_R^d d_R 
 \notag \\ & 
 + \frac{\sqrt{2}}{v} \overline{u}_L S_L^{u \dagger} \left[ 
   e^{- i \alpha} S_L^d M_D S_R^{d \dagger} G^{+} 
 +  \frac{1}{\cos 2 \beta}\left( 
 S_L^u M_U S_R^{u \dagger} 
 - \sin 2 \beta e^{- i \alpha} S_L^d M_D S_R^{d \dagger} \right) H^{+} \right] S_R^d d_R 
 \notag \\ &
 + \frac{\sqrt{2}}{v} \overline{d}_L S_L^{d \dagger} \left[ 
 - e^{  i \alpha} S_L^u M_U S_R^{u \dagger} G^{-} 
 - \frac{1}{\cos 2 \beta} \left( 
 S_L^d M_D S_R^{d \dagger} 
 - \sin 2 \beta e^{  i \alpha} S_L^u M_U S_R^{u \dagger} \right) H^{-} \right] S_R^u u_R 
 +  \textrm{h.c.}
 \notag \\ =\,& 
   \frac{\sqrt{2}}{v} \overline{u}_L \left[ 
 M_U h^{\ast} 
 + \frac{1}{\cos 2 \beta} \left( 
 V_L           M_D V_R^{\dagger} - \sin 2 \beta e^{  i \alpha} M_U \right) H^{0\ast} \right] u_R 
 \notag \\ & 
 + \frac{\sqrt{2}}{v} \overline{d}_L \left[ 
 M_D h 
 + \frac{1}{\cos 2 \beta} \left( 
 V_L^{\dagger} M_U V_R           - \sin 2 \beta e^{- i \alpha} M_D \right) H^0 \right] d_R 
 \notag \\ & 
 + \frac{\sqrt{2}}{v} \overline{u}_L \left[ 
   e^{- i \alpha} V_L           M_D G^{+} 
 + \frac{1}{\cos 2 \beta} \left( 
 M_U V_R 
 - \sin 2 \beta e^{- i \alpha} V_L           M_D \right) H^{+} \right] d_R 
 \notag \\ &
 + \frac{\sqrt{2}}{v} \overline{d}_L \left[ 
 - e^{  i \alpha} V_L^{\dagger} M_U G^{-} 
 - \frac{1}{\cos 2 \beta} \left( 
 M_D V_R^{\dagger} 
 - \sin 2 \beta e^{  i \alpha} V_L^{\dagger} M_U \right) H^{-} \right] u_R 
 + \textrm{h.c.}.
\end{align}
Therefore, the heavy Higgs interactions with quarks become
\begin{align}
 -\mathcal{L}_Y \simeq\,&
   \frac{\sqrt{2} m_{u_k}}{v \cos 2 \beta} 
 \overline{d}_{i} ( V_L^{\dagger} )_{ik} ( V_R )_{kj} P_R d_{j} H^0 
 + \frac{\sqrt{2} m_{u_k}}{v \cos 2 \beta} 
 \overline{d}_{i} ( V_R^{\dagger} )_{ik} ( V_L )_{kj} P_L d_{j} H^{0\ast} 
 \notag \\ & 
 + \frac{\sqrt{2} m_{u_k}}{v \cos 2 \beta} 
 \overline{u}_{k} (V_R)_{ki}           P_R d_{i} H^{+}  
 + \frac{\sqrt{2} m_{u_k}}{v \cos 2 \beta} 
 \overline{d}_{i} (V_R^{\dagger})_{ik} P_L u_{k} H^{-},
\end{align}
where the terms proportional to $\tan 2 \beta$ are dismissed, because $\tan 2 \beta = \mathcal{O}(m_b/m_t)$.
 
After integrating out the heavy charged Higgs boson, one obtains the effective operator,
\begin{align}
 \mathcal{L}_{\textrm{eff}} &\simeq 
   \frac{2 \sqrt{2} G_F}{\cos^2 2 \beta}
   \frac{m_t^2}{M^2_{H^{\pm}}}
 (V_R^{\dagger})_{i 3} (V_R)_{3 j}
 (\bar{d}_i P_L t )(\bar{t} P_R d_j) 
 \notag \\ &= 
 - \frac{  \sqrt{2} G_F}{\cos^2 2 \beta}
   \frac{m_t^2}{M^2_{H^{\pm}}}
 (V_R^{\dagger})_{i 3} (V_R)_{3 j}
 (\bar{t}_{\alpha} \gamma^{\mu} P_L t_{\beta})(\bar{d}_{i,\beta} \gamma_{\mu} P_R d_{j,\alpha}),
\end{align}
where $\alpha,\beta$ denote color indices.
By rearranging the colors, the Wilson coefficients become
\begin{align}
 (C_{qd}^{(8)})_{3 3 i j}^{\textrm{tree}}\big|_{q=u} &= 
 - \frac{2\sqrt{2} G_F}{\cos^2 2\beta} 
 \frac{m_t^2}{M^2_{H^{\pm}}}
 (V_R^{\dag})_{i 3} (V_R)_{3j},
 \\
 (C_{qd}^{(1)})_{3 3 i j}^{\textrm{tree}}\big|_{q=u} &= 
 \frac{1}{2 N_c} (C_{qd}^{(8)})_{33ij}\big|_{q=u}.
\end{align}
for $q=u$.
The Wilson coefficients for $q=d$ is generated by the heavy neutral Higgs exchange.
After integrating out the heavy neutral Higgs boson, one obtains
\begin{align}
 \mathcal{L}_{\textrm{eff}} &\simeq 
   \frac{2 \sqrt{2} G_F}{\cos^2 2 \beta}
   \frac{m_t^2}{M^2_{H^{0}}}
 ( V_R^{\dagger} )_{i3} ( V_L )_{3k}
 ( V_L^{\dagger} )_{l3} ( V_R )_{3j} 
 (\bar{d}_i P_L d_k)(\bar{d}_l P_R d_j) 
 \notag \\ &= 
 - \frac{  \sqrt{2} G_F}{\cos^2 2 \beta}
   \frac{m_t^2}{M^2_{H^{0}}}
 ( V_R^{\dagger} )_{i3} ( V_L )_{3k}
 ( V_L^{\dagger} )_{l3} ( V_R )_{3j} 
 (\bar{d}_{l,\alpha} \gamma^{\mu} P_L d_{k,\beta})(\bar{d}_{i,\beta} \gamma_{\mu} P_R d_{j,\alpha}).
\end{align}
In the mass eigenstate basis, the ${\rm SU(2)}_L$ quark double is shown as $q = (u_L, V_L d_L)^T$.
Thus, the Wilson coefficients for $q=d$ become
\begin{align}
 (C_{qd}^{(8)})_{3 3 i j}^{\textrm{tree}}\big|_{q=d} &= 
 - \frac{2\sqrt{2} G_F}{\cos^2 2\beta} 
 \frac{m_t^2}{M^2_{H}}
 (V_R^{\dag})_{i 3} (V_R)_{3j},
 \\
 (C_{qd}^{(1)})_{3 3 i j}^{\textrm{tree}}\big|_{q=d} &= 
 \frac{1}{2 N_c} (C_{qd}^{(8)})_{33ij}\big|_{q=d}.
\end{align}
Consequently, Eqs.~\eqref{eq:chH1} and \eqref{eq:chH2} are obtained.

\section{Loop functions}
\label{sec:loop}

The loop functions which are necessary for the $\Delta F=2$ transition amplitudes in the left-right model are summarized.
They are defined as
\begin{align}
 \mathcal{F}_A (x_i,x_j,\beta) =\,& 
   \left( 1 + \frac{x_i x_j \beta}{4} \right) \mathcal{I}_1 (x_i,x_j,\beta) 
 - \frac{ 1 + \beta }{4} \mathcal{I}_2 (x_i,x_j,\beta),
 \\
 \mathcal{F}_B (\tau_L,\tau_R) =\,&
 (\tau_L^2 + \tau_R^2 + 10 \tau_L \tau_R + 1)\, \mathcal{I}_3 (\tau_L,\tau_R)
 \notag \\ &
 + (\tau_L^2 + \tau_R^2 + 10 \tau_L \tau_R - 2\tau_L - 2\tau_R + 1)\, \mathcal{I}_4 (\tau_L,\tau_R),
 \\
 \mathcal{F}_C (\tau_i,\tau_j,\tau_L,\tau_R) =\,&
 2(\tau_L + \tau_R)\, \mathcal{I}_3 (\tau_L,\tau_R)
 - \left[
 \frac{ \tau_i \sqrt{\tau_L \tau_R} }{\tau_i - 4 \sqrt{\tau_L \tau_R}} 
 \mathcal{I}_5 (\tau_i,\tau_L,\tau_R) 
 + (i \to j) \right], \\
 \mathcal{F}_{D} (x_i,x_j,\tau_L) =\,&   
 x_i x_j  \mathcal{I}_1 \left( x_i, x_j, \tau_L \right) - \mathcal{I}_2 \left( x_i, x_j, \tau_L \right).
\end{align}
The functions, $\mathcal{I}_1$--$\mathcal{I}_5$, are denoted by the Passarino-Veltman functions as~\cite{Passarino:1978jh}
\begin{align}
 \mathcal{I}_1 (x_i,x_j,\beta) &= 
 - M_{W_L}^2 M_{W_R}^2 D_{0} (0, 0, 0, 0; 0, 0;m_{u_i},m_{u_j},M_{W_L},M_{W_R}),
 \\
 \mathcal{I}_2 (x_i,x_j,\beta) &= 
 - 4 M_{W_R}^2 D_{00} (0, 0, 0, 0; 0, 0;m_{u_i},m_{u_j},M_{W_L},M_{W_R}),
 \\
 \mathcal{I}_3 (\tau_L,\tau_R) &= 
 B_{0} (0;M_{W_L},M_{W_R})-{\rm Re}[B_{0} (M_{H}^2;M_{W_L},M_{W_R})],
 \\
 \mathcal{I}_4 (\tau_L,\tau_R) &= 
 M_H^2 \sum_{n=0}^2
 {\rm Re}\left[ 
 C_{n} (M_{H}^2,0,M_{H}^2;M_{W_L},M_{W_R},M_{W_R})\right],
 \\
 \mathcal{I}_5 (\tau_i,\tau_L,\tau_R) &= 
 M_H^2 \bigg\{ C_{0} (0,0,0;M_{W_L},m_{u_i},M_{W_R})
 \notag \\ & \quad\quad\quad
 - {\rm Re}\left[
 C_{0} \left(\frac{M_{H}^2}{4},\frac{M_{H}^2}{4} ,M_{H}^2;M_{W_L},m_{u_i},M_{W_R}\right)
 \right] \bigg\},
\end{align}
where we follow the notation of Refs.~\cite{Patel:2015tea,Patel:2016fam}.
The absorptive parts in the loop functions are discarded~\cite{Bertolini:2014sua}.
We also obtain the following analytical formulae:
\begin{align}
 \mathcal{I}_1 (x_i,x_j,\beta) &= 
   \frac{x_i \ln x_i}{(1-x_i)(1-x_i \beta)(x_i-x_j)}
 + (i \leftrightarrow j)
 - \frac{\beta \ln \beta}{(1-\beta)(1-x_i \beta)(1-x_j \beta)},
 \\
 \mathcal{I}_2 (x_i,x_j,\beta) &=
   \frac{x_i^2 \ln x_i}{(1-x_i)(1-x_i \beta)(x_i-x_j)}
 + (i \leftrightarrow j)
 - \frac{\ln \beta}{(1-\beta)(1-x_i \beta)(1-x_j \beta)},
 \\
 \mathcal{I}_3 (\tau_L,\tau_R) &=
  - 1 + \frac{1}{2}\left[
   \tau_L - \tau_R 
 - \frac{\tau_L + \tau_R}{\tau_L - \tau_R} \right]
 \ln \frac{\tau_L}{\tau_R}
 \\ &
 - \frac{ \sqrt{(1 - \tau_L - \tau_R)^2 - 4 \tau_L \tau_R} }{2} 
 \ln \frac{1 - \tau_L - \tau_R - \sqrt{(1 - \tau_L - \tau_R)^2 - 4 \tau_L \tau_R}}
          {1 - \tau_L - \tau_R + \sqrt{(1 - \tau_L - \tau_R)^2 - 4 \tau_L \tau_R}},
 \notag \\
 \mathcal{I}_4 (\tau_L,\tau_R) &=
 1 - \frac{\tau_L - \tau_R}{2} \ln \frac{\tau_L}{\tau_R}
 \\ &
 + \frac{ (\tau_L - \tau_R)^2 - (\tau_L + \tau_R) }
        { 2 \sqrt{ (1 - \tau_L - \tau_R)^2 - 4 \tau_L \tau_R } }
 \ln \frac{1 - \tau_L - \tau_R - \sqrt{(1 - \tau_L - \tau_R)^2 - 4 \tau_L \tau_R}}
          {1 - \tau_L - \tau_R + \sqrt{(1 - \tau_L - \tau_R)^2 - 4 \tau_L \tau_R}},
 \notag \\
 \mathcal{I}_5 (\tau_i,\tau_L,\tau_R) &=
 \frac{ \tau_i (\tau_R -\tau_L) \ln \tau_i 
      + \tau_L (\tau_i -\tau_R) \ln \tau_L 
      + \tau_R (\tau_L -\tau_i) \ln \tau_R }
      { (\tau_R - \tau_L)(\tau_L - \tau_i)(\tau_i - \tau_R) }
 \notag \\ &
 - {\rm Re} \bigg\{
 \ln \frac{\tau_L\tau_R}{\tau_i^2} + \frac{1}{4\tau_i-2\tau_L-2\tau_R+1}
 \notag \\ & \quad
 \times \bigg[ 
 8\kappa\left(\frac{1}{4},\tau_i,\tau_L\right) 
 \ln \frac{\kappa\left(\frac{1}{4},\tau_i,\tau_L\right) + \tau_i + \tau_L - \frac{1}{4}}
          {2 \sqrt{\tau_i\tau_L}} 
 + (L \to R)
 \notag \\ & \quad\quad
 -4\kappa(1,\tau_L,\tau_R) 
 \ln \frac{\kappa(1,\tau_L,\tau_R) + \tau_L + \tau_R - 1}{2 \sqrt{\tau_L\tau_R}} \bigg]
 \bigg\},
 \label{eq:I5}
\end{align}
with 
\begin{align}
 \kappa(x,y,z) = \sqrt{x^2 + y^2 + z^2 - 2 (xy + yz + zx)}.
\end{align}
When the relation, $m_{u_i}^2$, $M_{W_L}^2 \ll M_{W_R}^2 \ll M_H^2$, are satisfied, one can use the following approximations:
\begin{align}
 \mathcal{I}_3 (\tau_L,\tau_R) &\simeq 
 -1 + \left( 1 - \tau_R \right) \ln \left( \frac{1}{\tau_R} -1 \right) 
 \approx -1 - \ln \tau_R,\\
 \mathcal{I}_4 (\tau_L,\tau_R) &\simeq 
 1 + \tau_R  \ln \left( \frac{1}{\tau_R} -1 \right) \approx 1, \\
 \mathcal{I}_5 (\tau_i,\tau_L,\tau_R) &\simeq 
 \frac{\tau_i \ln \left( \tau_i /\tau_R\right) 
 - \tau_L \ln \left( \tau_L /\tau_R\right)}{\tau_R (\tau_i - \tau_L)}, 
\end{align}
which are consistent with Ref.~\cite{Bertolini:2014sua}.
Numerically, the second term of  $\mathcal{I}_5$ in Eq.~\eqref{eq:I5}, Re$\{\cdots \}$, is much smaller than the first term for $M_H \gg M_{W_R}$.

\section{Double penguin contributions}
\label{sec:DoublePenguin}

In this section, we apply the one-loop matching conditions in Sec.~\ref{sec:formula} to double-penguin diagrams, where $\Delta F=2$ processes are generated by exchanging the SM gauge bosons with FC interactions. 
When vector bosons of the unbroken gauge symmetries, i.e., those of ${\rm SU(3)}_C$ and ${\rm U(1)}_{em}$ in the SM, are exchanged, such double-penguin contributions should vanish because of the gauge invariance. 
In fact, form factors of their FC penguin vertices should be proportional to $q^2$, i.e., vanish in the limit of $q^2 \to 0$ for the gauge invariance, where $q$ is the momentum transfer.
Then, $\Delta F=2$ double-penguin diagrams depend on $q^4 \times 1/q^2$, where $1/q^2$ represents the propagator of the unbroken gauge boson. 
Hence, they disappear in the limit of $q^2 \to 0$. 

In our formula, this gauge invariance is confirmed by observing the cancellations among the Wilson coefficients.
Once $\Delta F=1$ operators (and $\Delta F=2$ ones if necessary) are generated by the penguin diagrams at the NP scale, we will see that $\Delta F=2$ contributions cancel out below the EWSB scale, if the diagrams are mediated by the gauge bosons of the unbroken gauge symmetries. 
Here, the one-loop matching conditions are necessary.
These results justify our one-loop matching conditions in Sec.~\ref{sec:formula}.  

We will focus on the double-penguin diagrams with exchanging the gauge bosons associated with the unbroken gauge symmetries.
At the NP scale, penguin-type $\Delta F=1$ contributions are generated by exchanging them. 
The effective Lagrangian from the massless $B$, $W^3$ and gluon can be written as
\begin{align}
 \mathcal{L}^B =& 
 \frac{\alpha}{4\pi c_W^2} 
 \left(
   C_{L,ij}^B \bar{d}_i \gamma^{\mu} P_L d_j 
 + C_{R,ij}^B \bar{d}_i \gamma^{\mu} P_R d_j \right)
 \left(
   Y_{u_L} \bar{u}_k \gamma_{\mu} P_L u_k 
 + Y_{u_R} \bar{u}_k \gamma_{\mu} P_R u_k \right)
 \notag \\
 &+ \frac{\alpha}{4\pi c_W^2} 
 \left(
   C_{L,ij}^B \bar{d}_i \gamma^{\mu} P_L d_j 
 + C_{R,ij}^B \bar{d}_i \gamma^{\mu} P_R d_j \right)
 \left(
   Y_{d_L} \bar{d}_k \gamma_{\mu} P_L d_k 
 + Y_{d_R} \bar{d}_k \gamma_{\mu} P_R d_k \right)
 \notag \\
  =&
 \frac{\alpha}{4\pi}
 \left(
   C_{L,ij}^B \bar{d}_i \gamma^{\mu} P_L d_j
 + C_{R,ij}^B \bar{d}_i \gamma^{\mu} P_R d_j \right)
 Q_{u} \bar{u}_k \gamma_{\mu} u_k
 \notag \\ &
 - \frac{\alpha_Z}{4\pi}s_W^2 
 \left(
   C_{L,ij}^B \bar{d}_i \gamma^{\mu} P_L d_j
 + C_{R,ij}^B \bar{d}_i \gamma^{\mu} P_R d_j \right)
 \left( I^3_{u} - Q_{u} s_W^2 \right) 
 \bar{u}_k \gamma_{\mu} u_k\notag
 \\
  &+\frac{\alpha}{4\pi}
 \left(
   C_{L,ij}^B \bar{d}_i \gamma^{\mu} P_L d_j
 + C_{R,ij}^B \bar{d}_i \gamma^{\mu} P_R d_j \right)
 Q_{d} \bar{d}_k \gamma_{\mu} d_k
 \notag \\ &
 - \frac{\alpha_Z}{4\pi}s_W^2 
 \left(
   C_{L,ij}^B \bar{d}_i \gamma^{\mu} P_L d_j
 + C_{R,ij}^B \bar{d}_i \gamma^{\mu} P_R d_j \right)
 \left( I^3_{d} - Q_{d} s_W^2 \right) 
 \bar{d}_k \gamma_{\mu} d_k,
 \\
 \mathcal{L}^{W^3} =&
 \frac{\alpha}{4\pi s_W^2} C_{L,ij}^{W^3} 
 \left( \bar{d}_i \gamma^{\mu} P_L d_j \right) 
 \left( I^3_{u} \bar{u}_k \gamma_{\mu} P_L u_k \right)
+ \frac{\alpha }{4\pi s_W^2} C_{L,ij}^{W^3} 
 \left( \bar{d}_i \gamma^{\mu} P_L d_j \right) 
 \left( I^3_{d} \bar{d}_k \gamma_{\mu} P_L d_k \right)
 \notag \\
  =&
 \frac{\alpha}{4\pi} C_{L,ij}^{W^3} 
 \left( \bar{d}_i \gamma^{\mu} P_L d_j \right) 
 Q_{u} \bar{u}_k \gamma_{\mu} u_k + \frac{\alpha_Z}{4\pi} c_W^2 C_{L,ij}^{W^3} 
 \left( \bar{d}_i \gamma^{\mu} P_L d_j \right) 
 \left( I^3_{u} - Q_{u} s_W^2 \right) 
 \bar{u}_k \gamma_{\mu} u_k\notag
 \\
  &+\frac{\alpha}{4\pi} C_{L,ij}^{W^3} 
 \left( \bar{d}_i \gamma^{\mu} P_L d_j \right) 
 Q_{d} \bar{d}_k \gamma_{\mu} d_k
 + \frac{\alpha_Z}{4\pi} c_W^2 C_{L,ij}^{W^3} 
 \left( \bar{d}_i \gamma^{\mu} P_L d_j \right) 
 \left( I^3_{d} - Q_{d} s_W^2 \right) 
 \bar{d}_k \gamma_{\mu} d_k,
 \\ 
 \mathcal{L}^g =&
 \frac{\alpha_s}{4\pi}
 \left(
   C_{L,ij}^g \bar{d}_i \gamma^{\mu} P_L T^A d_j 
 + C_{R,ij}^g \bar{d}_i \gamma^{\mu} P_R T^A d_j \right)
 \left(
   \bar{u}_k \gamma_{\mu} T^A u_k 
 + \bar{d}_k \gamma_{\mu} T^A d_k \right),
\end{align}
where $Y_f$ is the hypercharge, $I^3_f$ the ${\rm SU(2)}_L$ charge, and $Q_{f}$ the ${\rm U(1)}_{em}$ charge. 
Also, $\alpha_s$ and $\alpha$ are the gauge couplings of ${\rm SU(3)}_C$  and ${\rm U(1)}_{em}$, respectively. 
The coefficients, $C_{c,ij}^{V}$ $(V = B,W^3,g$ and $c =L,R)$, are generated by integrating out the NP particles.
In the second lines of $\mathcal{L}^B$ and $\mathcal{L}^{W^3}$, the effective Lagrangians are divided into the would-be $\gamma$- and $Z$-penguin contributions, which are proportional to $\alpha\, Q_f$ and $\alpha_Z \,(I_f^3  -Q_f s_W^2)$, respectively. 
Here, $\alpha_Z = \alpha/(c_W^2 s_W^2)$.
In terms of the SMEFT operators, the above operators are represented as
\begin{align}
 (\mathcal{C}_{qq}^{(1)})_{ijkk} =&
 (\mathcal{C}_{qq}^{(1)})_{kkij} 
 \notag \\ =& 
 - \frac{\alpha_s}{16 N_c\pi}C_{L,ij}^g
 + \frac{\alpha}{8 \pi}
 Y_q 
 \left( C^B_{L,ij} + C^{W^3}_{L,ij} \right)
 + \frac{\alpha_Z}{8 \pi} s_W^2 
 Y_q 
 \left( s_W^2 C^B_{L,ij} - c_W^2 C^{W^3}_{L,ij} \right),
 \label{SMEFTD4}
 \\
 (\mathcal{C}_{qq}^{(1)})_{ikkj} =& 
 (\mathcal{C}_{qq}^{(1)})_{kjik} = 
 \frac{\alpha_s}{32N_c \pi} (N_c-2) C_{L,ij}^g,
 \\
 (\mathcal{C}_{qq}^{(3)})_{ijkk} =&
 (\mathcal{C}_{qq}^{(3)})_{kkij} 
 \notag \\ = &
 - \frac{\alpha}{16\pi}
 \left( C^B_{L,ij} + C^{W^3}_{L,ij} \right)
 + \frac{\alpha_Z}{16\pi}c_W^2
 \left( s_W^2 C^B_{L,ij} - c_W^2 C^{W^3}_{L,ij} \right),
 \\ 
 (\mathcal{C}_{qq}^{(3)})_{ikkj} =& 
 (\mathcal{C}_{qq}^{(3)})_{kjik} =
 \frac{\alpha_s}{32\pi} C_{L,ij}^g,
 \\
 (\mathcal{C}_{ud}^{(1)})_{kkij} =&
 \frac{\alpha}{4\pi} Q_u C^B_{R,ij}
 - \frac{\alpha_Z}{4\pi} s_W^2 C^B_{R,ij} 
 \left( -s_W^2 Q_u \right),
 \\
 (\mathcal{C}_{ud}^{(8)})_{kkij} =&
 \frac{\alpha_s}{4\pi} C_{R,ij}^g,
 \\
 (\mathcal{C}_{qu}^{(1)})_{ijkk} =&
 \frac{\alpha}{4\pi} Q_u 
 \left( C^B_{L,ij} + C^{W^3}_{L,ij} \right)
 - \frac{\alpha_Z}{4\pi}
 \left( s_W^2 C^B_{L,ij} - c_W^2 C^{W^3}_{L,ij} \right)
 \left( -s_W^2 Q_u \right),
 \\
 (\mathcal{C}_{qu}^{(8)})_{ijkk} =&
 \frac{\alpha_s}{4\pi} C_{L,ij}^g,
 \\
 (\mathcal{C}_{qd}^{(1)})_{kkij} =&
 \frac{\alpha}{4\pi} Q_{q} C^B_{R,ij} 
 - \frac{\alpha_Z}{4\pi} s_W^2 C^B_{R,ij}
 \left( I^3_{q} - s_W^2 Q_{q} \right),
 \\
 (\mathcal{C}_{qd}^{(8)})_{ijkk} =&
 \frac{\alpha_s}{4\pi} C_{L,ij}^g,
 \\
 (\mathcal{C}_{qd}^{(8)})_{kkij} =&
 \frac{\alpha_s}{4\pi} C_{R,ij}^g.
 \label{SMEFTD14}
\end{align}

\begin{figure}[t]
\begin{center}
\includegraphics[width=0.3\textwidth, bb= 0 0 338 210]{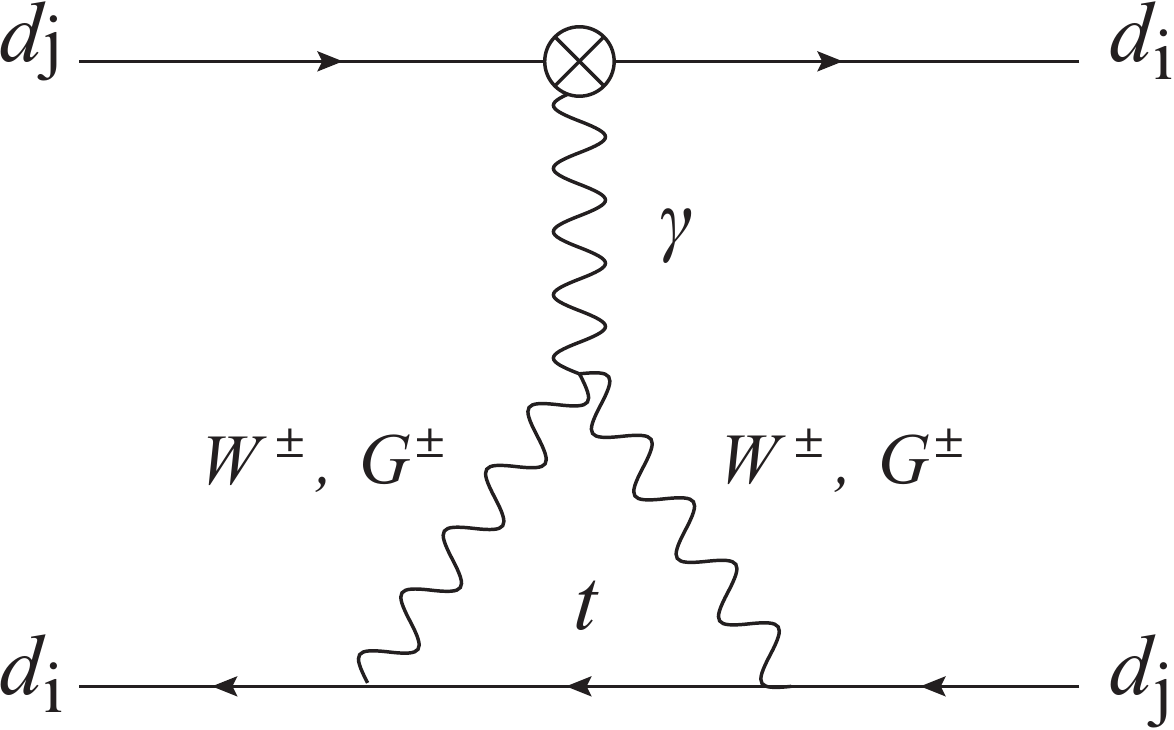}
\caption{
Feynman diagram for the one-loop contribution to $\Delta F = 2$ operators  at the NP scale.
}
\label{fig:photonpenguin}
\end{center}
\end{figure}

In addition, one has to include $\Delta F=2$ contributions which come from the diagram in  Fig.~\ref{fig:photonpenguin}. 
They are generated at the NP scale. 
The $d_i \to d_j$ transitions are induced by the penguin vertices of the NP contribution in one side and those of the SM contribution in another side, where the up-type quarks, especially the top quark, and the $W$ boson are exchanged. 
The Wilson coefficients of the SMEFT  operators are represented as 
\begin{align}
(\mathcal{C}_{qq}^{(1)})_{ijij}=&(\mathcal{C}_{qq}^{(3)})_{ijij}\notag
\\
=&-\frac{\alpha^2 \lambda_t^{ij}}{8\pi^2 s_W^2}\left(C^B_{L,ij}+C^{W^3}_{L,ij} \right)\left\{Q_{G^+}\frac{x_t}{8}L(x_t,\mu_W)+Q_{G^+}\frac{1}{4}M(x_t)-\frac{1}{8}\left[2-6 L(x_t,\mu_W)\right] \right\}
 \notag
\\
&+\frac{\alpha\alpha_Z \lambda_t^{ij}}{8\pi^2 s_W^2}\left(s_W^2 C^B_{L,ij}-c_W^2 C^{W^3}_{L,ij} \right)\bigg\{(I^{G^+}-s_W^2 Q^{G^+})\frac{x_t}{8}L(x_t,\mu_W)\notag
\\
&+(-s_W^2Q_{G^+})\frac{1}{4}M(x_t) -\frac{1}{8}c_W^2\left[2-6 L(x_t,\mu_W)\right] \bigg\}
\\
 =&
 -\frac{\alpha^2 \lambda_t^{ij}}{8\pi^2 s_W^2}
 \left( C_{L,ij}^B + C_{L,ij}^{W^3} \right) K(x_t,\mu_W)\notag
 \\
 &-\frac{\alpha\alpha_Z \lambda_t^{ij}}{8\pi^2 s_W^2}
 \left( s_W^2C_{L,ij}^B - c_W^2 C_{L,ij}^{W^3} \right) \left[ \frac{x_t}{16}L(x_t,\mu_W)+\frac{1}{4}M(x_t)-c_W^2 K(x_t,\mu_W) \right] , 
 \label{eq:phmat1}
 \\
 (\mathcal{C}_{qd}^{(1)})_{ijij}=&-\frac{\alpha^2 \lambda_t^{ij}}{4\pi^2 s_W^2}C^B_{R,ij}\left\{  Q_{G^+}\frac{x_t}{8}L(x_t,\mu_W)+Q_{G^+}\frac{1}{4}M(x_t)-\frac{1}{8}\left[2-6 L(x_t,\mu_W)\right] \right\}
 \notag
\\
&+\frac{\alpha\alpha_Z \lambda_t^{ij}}{4\pi^2 s_W^2}\left(s_W^2 C^B_{R,ij} \right)\bigg\{ (I^{G^+}-s_W^2 Q^{G^+})\frac{x_t}{8}L(x_t,\mu_W)+(-s_W^2Q_{G^+})\frac{1}{4}M(x_t)\notag
\\
&-\frac{1}{8}c_W^2\left[ 2-6 L(x_t,\mu_W)\right] \bigg\}
\\
 =&
 -\frac{\alpha^2 \lambda_t^{ij}}{4\pi^2 s_W^2}
 C_{R,ij}^B K(x_t,\mu_W)\notag
 \\
 &-\frac{\alpha\alpha_Z \lambda_t^{ij}}{4\pi^2 }
 C_{R,ij}^B  \left[\frac{x_t}{16}L(x_t,\mu_W)+\frac{1}{4}M(x_t)-c_W^2K(x_t,\mu_W) \right] ,
 \label{eq:phmat2}
\end{align}
where $Q_{G^+}=1$ and $I^{G^+}=1/2$.
Here, the GIM mechanism is used to reduce the results, and the loop functions are given as
\begin{align}
 L(x,\mu) &= \ln \frac{\mu}{M_W} + \frac{3x -1}{4(x-1)} - \frac{x^2 \ln x}{2(1-x)^2},
 \\
 M(x) &= \frac{x}{1-x} + \frac{x^2 \ln x}{(1-x)^2}.
\end{align}

For the gluon double-penguin contributions, one obtains the low-scale $\Delta F=2$ operators through the one-loop matching conditions, Eqs.~\eqref{eq:EWSBmatchingLoop1}--\eqref{eq:EWSBmatchingLoop3},  from the SMEFT $\Delta F=1$ operators in Eqs.~\eqref{SMEFTD4}--\eqref{SMEFTD14} as
\begin{align}
 (C_1)_{ij} &= (C_1)_{ij}^{\textrm{1--loop}}\notag \\
 & = 
 -\frac{\alpha\lambda_t^{ij}}{4\pi s_W^2}
 \frac{\alpha_s}{4\pi} \frac{2(N_c-1)}{N_c}
 C_{L,ij}^g
 \left[ I_1 (x_t,\mu_W) + 2 J(x_t) - K(x_t,\mu_W) \right] \nonumber \\ 
 &= 0,
 \\
 ({C}_4)_{ij} &= (C_4)_{ij}^{\textrm{1--loop}}\notag \\
 & = 
 \frac{\alpha\lambda_t^{ij}}{\pi s_W^2}
 \frac{\alpha_s}{4\pi} 
 C^g_{R,ij}
 \left[ I_1(x_t,\mu_W) + 2 J(x_t) - K(x_t,\mu_W) \right] \nonumber \\ 
 & = 0,
 \\
 ({C}_5)_{ij} &= (C_5)_{ij}^{\textrm{1--loop}}\notag \\
 & = 
 -\frac{\alpha\lambda_t^{ij}}{\pi s_W^2}
 \frac{\alpha_s}{4\pi} \frac{1}{N_c}
 C_{R,ij}^g 
 \left[ I_1(x_t,\mu_W) + 2 J(x_t) - K(x_t,\mu_W) \right] \nonumber \\ 
 &= 0.
\end{align}
Since all these Wilson coefficients are proportional to the function, $I_1(x_t,\mu_W) + 2 J(x_t) - K(x_t,\mu_W)$, which is identical to zero, there are no contributions to the $\Delta F =2$ operators.  
Hence, the gluon double-penguin contributions vanish, as expected from the gauge invariance. 

Next, for the $\gamma$ double-penguin contributions, the low-scale $\Delta F=2$ operators are generated from the SMEFT $\Delta F=2$ operators in Eqs.~\eqref{eq:phmat1} and \eqref{eq:phmat2} through the tree-level matching as well as the $\Delta F=1$ ones in Eqs.~\eqref{SMEFTD4}--\eqref{SMEFTD14} through the one-loop matching conditions, Eqs.~\eqref{eq:EWSBmatchingLoop1}--\eqref{eq:EWSBmatchingLoop3}. 
In total, the low-scale $\Delta F=2$ coefficients are 
\begin{align}
 (C_1)_{ij} =&  (C_1)_{ij}^{\rm tree} +(C_1)_{ij}^{\textrm{1--loop}} \notag \\
=&   
\frac{\alpha \lambda_t^{ij}}{\pi s_W^2}
 \frac{\alpha}{4\pi}
 \left( C_{L,ij}^B + C_{L,ij}^{W^3} \right)  K(x_t,\mu_W) \notag \\
&-\frac{\alpha \lambda_t^{ij}}{\pi s_W^2}
 \frac{\alpha}{4\pi}
 \left( C_{L,ij}^B + C_{L,ij}^{W^3} \right)
 \Big\{  Q_u [ I_1(x_t,\mu_W) + 2 J(x_t) ] - Q_d K(x_t,\mu_W) \Big\} \notag \\
 =& 0,
 \\
 (C_5)_{ij} =&  (C_5)_{ij}^{\rm tree} +(C_5)_{ij}^{\textrm{1--loop}} \notag \\
 =& 
  - \frac{2\alpha\lambda_t^{ij}}{\pi s_W^2}
 \frac{\alpha}{4\pi} 
 C_{R,ij}^B  K(x_t,\mu_W) \notag \\
& +
 \frac{2\alpha\lambda_t^{ij}}{\pi s_W^2}
 \frac{\alpha}{4\pi} 
 C_{R,ij}^B
 \Big\{     Q_u [ I_1(x_t,\mu_W) + 2 J(x_t) ] - Q_d K(x_t,\mu_W)  \Big\} \notag \\
  =& 0,
\end{align}
and other Wilson coefficients do not receive contributions. 
It is noticed that $(C_1)_{ij}$ and $ (C_5)_{ij}$ are proportional to the function which is identical to zero, because $Q_d = Q_u -  1$. 
Hence, the $\gamma$ double-penguin contributions also vanish, as expected from the gauge invariance, and it guarantees our one-loop matching conditions.\footnote{
 The $Z$ double-penguin contributions also vanish in the limit of the gauge invariance of the ${\rm SU(2)}_L \times {\rm U(1)}_{Y}$ symmetry. 
 Non-zero contributions due to the ${\rm SU(2)}_L \times {\rm U(1)}_{Y}$ breaking are encoded into the $\Delta F = 1$ effective operators, $(\mathcal{O}_{Hq}^{(1)})_{ij}, (\mathcal{O}_{Hq}^{(3)})_{ij}$ and $(\mathcal{O}_{Hd})_{ij}$, in a gauge-invariant manner~\cite{Endo:2016tnu, Bobeth:2017xry} (see also Ref.~\cite{Endo:2017ums} for a supersymmetric study).
}


\end{document}